\documentclass[a4paper,twocolumn,11pt,accepted=2022-12-13]{quantumarticle}
\pdfoutput=1
\usepackage{mathtools,amssymb,amsthm,bm,bbm,xcolor,mathdots,stmaryrd}
\usepackage{hyperref}
\hypersetup{
    colorlinks=true,
    linkcolor=blue,
    filecolor=red,      
    urlcolor=teal,citecolor=purple
}

\usepackage[utf8]{inputenc}
\usepackage[english]{babel}
\usepackage[T1]{fontenc}

\usepackage{bbold}
\usepackage{amsmath, dsfont}
\usepackage{amsfonts}
\usepackage{amsthm}
\usepackage[normalem]{ulem}
\usepackage{tikz}
\usepackage{ifthen}
\usepackage{stmaryrd}
\usetikzlibrary{tikzmark}
\usepackage{upgreek}
\usepackage{braket}
\usepackage{graphicx}
\usepackage[shortlabels]{enumitem}
\usepackage{tikz-cd}
\usepackage{systeme}
\usepackage{xparse}
\usepackage{mathrsfs}
\usepackage{wrapfig}

\DeclareMathOperator{\Tr}{Tr}

\usepackage[numbers]{natbib}
\begin{document}
\title{Quantum speed limits on operator flows and correlation functions}

\author{Nicoletta Carabba}
\affiliation{Department  of  Physics  and  Materials  Science,  University  of  Luxembourg,  L-1511  Luxembourg, G. D.  Luxembourg}
\author{Niklas H{\"o}rnedal}
\affiliation{Department  of  Physics  and  Materials  Science,  University  of  Luxembourg,  L-1511  Luxembourg, G. D.  Luxembourg}
\affiliation{Fysikum, Stockholms Universitet, 106 91 Stockholm, Sweden}
\author{Adolfo del Campo}
\affiliation{Department  of  Physics  and  Materials  Science,  University  of  Luxembourg,  L-1511  Luxembourg, G. D.  Luxembourg}
\affiliation{Donostia International Physics Center,  E-20018 San Sebasti\'an, Spain}

\begin{abstract}
Quantum speed limits (QSLs) identify fundamental time scales of physical processes by providing lower bounds on the rate of change of a quantum state or the expectation value of an observable.
We introduce a generalization of QSL for unitary operator flows, which are ubiquitous in physics and relevant for applications in both the quantum and classical domains. We derive two types of QSLs and assess the existence of a crossover between them, that we illustrate with a qubit and a random matrix Hamiltonian, as canonical examples. We further apply our results to the time evolution of autocorrelation functions, obtaining computable constraints on the linear dynamical response of quantum systems out of equilibrium and the quantum Fisher information governing the precision in quantum parameter estimation.
\end{abstract}

\maketitle

\section{Introduction}

Unraveling the fundamental time scale of a physical process is crucial in many theoretical and experimental scenarios.  Quantum speed limits (QSLs) constitute a set of fundamental results in quantum physics that bound the minimum time for a physical process to happen. Although their initial formulation \cite{Mandelstam45,Margolus98} was restricted to  unitary evolution between two pure quantum states, by now they have been generalized to the case of arbitrary mixed states \cite{Uhlmann1992,Campaioli2018}, driven Hamiltonians \cite{Anandan1990,Uhlmann1992,Deffner2013JPA,Okuyama18}, open quantum dynamics \cite{Taddei2013,Delcampo2013,Deffner2013PRL,Campaioli2019}, continuous quantum measurements \cite{GarciaPintos19}, and classical processes \cite{Shanahan2018,Okuyama2018PRL,Shiraishi18}. Their applications are thus manifold and range over various branches of physics  \cite{Deffner2017}. Their use is prominent in quantum technologies, including quantum computation \cite{Lloyd2000,Lloyd2002} and quantum metrology \cite{Giovannetti2011,Beau17}. 
QSL are known to limit the performance of quantum control algorithms \cite{Caneva2009,Hegerfeldt13}, and play a key role in shortcuts to adiabaticity by counterdiabatic driving both in isolated \cite{Funo17,Campbell2017} and open quantum systems \cite{Alipour20,Funo21}.
They have also been applied to many-body physics, in which care is needed given the orthogonality catastrophe \cite{Bukov2019,Suzuki20,Delcampo21,Hamazaki2021,Hamazaki2022}. 
In addition, some formulations of QSLs have transcended the notion of quantum state distinguishability and focused on the rate of change of other quantities such as quantum coherence \cite{Jing16,Marvian16,Mohan22}, and more generally, quantum resources \cite{Campaioli22}. 

Bounds to the pace of evolution naturally have important implications in nonequilibrium thermodynamics. In this context, the connection between QSLs  and thermodynamic uncertainty relations (TURs) \cite{Gingrich2016,Hasegawa2021} has been explored in \cite{Nicholson2020,Vo20}, identifying  QSL-inspired constraints on non-equilibrium fluctuations. Recently, a closely related work \cite{Garcia2022}, following the initial spirit of \cite{Mandelstam45}, established a generalized QSL on the evolution of observables, in particular, of their expectation values, under arbitrary dynamics. The notion of the speed of an observable expectation value was also considered in \cite{Hamazaki2021,Hamazaki2022,Mohan21}.

A natural question then arises: is it possible to formulate QSLs directly on the operator flow, rather than on the time-dependent expectation value of an observable? An operator flow describes the evolution of an operator under some given dynamics. In the unitary case, the evolution of the operator results from its conjugation by a unitary. Unitary operator flows are ubiquitous in physics, ranging from the quantum evolution of an observable in the Heisenberg picture to the Lax pair flow in classical integrable systems \cite{Perelomov1989integrable} and continuous renormalization group flows, such as the one proposed by Wegner \cite{Wegner2001}. Preceding efforts in this direction have focused on the preparation of unitary \cite{Poggi2019} and non-unitary \cite{Uzdin2013} quantum operations in the context of quantum control. Moreover, a Liouville space formulation of the QSL on the evolution of density matrices was introduced in \cite{Uzdin2016}. 
However, a framework to describe general operator flows, for general operators, is still lacking. Constraining the speed of these flows would naturally prove useful also in characterizing operator growth \cite{Keyserlingk2018,Khemani2018,Nahum2018,Gopalakrishnan2018,Rakovsky2018}, which has attracted increasing attention as a powerful approach to describe the buildup of complexity in quantum systems, in parallel to the earlier notion of quantum state complexity \cite{Susskind2016}, that is also subject to QSLs \cite{Brown16,Brown16prd,Chapman18,Molina-Vilaplana18}. In this context, a QSL-like uncertainty bound has recently been introduced \cite{Hornedal2022}  on the growth of a particular notion of operator complexity, known as Krylov complexity \cite{Parker2019,Barbon2019,Rabinovici2021,Caputa2021}.

In this manuscript, we introduce two notions of QSL for operator flows. Although we focus on the unitary time evolution in the Heisenberg picture, our results can be applied to a wider class of unitary flows, including some examples of inhomogeneous flows. Moreover, they do not require the flowing operator to be Hermitian. In addition, we introduce QSLs for the two-point autocorrelation functions, which play a key role in many-body physics, determine the linear response \cite{Kubo1957} and transport properties of a quantum system, and characterize the operator growth in Krylov space. The exact computation of correlation functions is generally a challenging task, requiring solving the dynamics, while our bounds are easy to compute provided that the Hamiltonian and the initial operator are known. Our results show the presence of a universal crossover between an initial regime in which the dynamics follows a Mandelstam-Tamm (MT) \cite{Mandelstam45} type of QSL and a second one in which a Margolus-Levitin (ML) \cite{Margolus98} type of QSL yields a more accurate description. We illustrate this crossover, analogous to that recently observed experimentally for quantum state evolution \cite{Ness2021}, in the case of a two-level system and of a random matrix Hamiltonian. Furthermore, our results provide easily computable constraints on the non-equilibrium dynamical response of arbitrary systems to an external perturbation within linear response theory. Finally, exploiting the relation between the dynamical susceptibility and the quantum Fisher information \cite{Hauke2016}, we upper bound this quantity with an easily computable correlation function.

\section{Quantum speed limits for operators}

The paradigmatic example of operator flow in quantum mechanics is the unitary flow $\dot{O}_t=\frac{i}{\hbar}[H,O_t],$ describing the time evolution of an initial operator $O_0$ generated by the Liouvillian {$\mathbb{L}[\cdot]= \frac{1}{\hbar}[H,\cdot]$. The unitary character of the flow is manifest in the formal solution  $O_t=U_t^\dag O_0U_t=\exp(it\mathbb{L})O_0$}, where $U_t=\exp(-itH/\hbar)$ is the unitary time-evolution operator associated with a time-independent Hamiltonian $H$. It is convenient to visualize this flow upon vectorizing operators in Liouville space \cite{Uzdin2016}, as it allows a direct analogy between operator flow and quantum state evolution. Therefore, let us represent a general bounded operator $A=\sum_{i,j}A_{ij}\ket{i}\bra{j}$, where $\{\ket{i}\}$ can be any basis of the Hilbert space, as a normalized vector
\begin{equation}\label{vector}
	\ket{A}=\frac{1}{\|A\|}\sum_{i,j}A_{ij}\ket{i}\otimes\ket{j},
\end{equation} 
where $\|A\|=\sqrt{\langle A,A\rangle}$ is the Hilbert-Schmidt  norm associated with the Hilbert-Schmidt inner product $\braket{A,B}=\Tr[A^\dagger B]$. In Liouville space, the latter is proportional to the standard scalar product over $\mathbb{C}^{d^2}$, being $\langle A,B\rangle=\|A\|\|B\|\braket{A|B}$. In the isomorphic real vector space $\mathbb{R}^{2d^2}$, the real part $\operatorname{Re} \braket{A,B}$ defines the standard Euclidean inner product. {Let us stress that, despite the formal similarity, vectorized operators and physical states are fundamentally different. A quantum state $|\psi\rangle$, being a  ray in the projective Hilbert space,  is defined up to a global phase, i.e.,  $\exp(i\theta)|\psi\rangle$ with $\theta \in \mathbb{R}$ is equivalent to $|\psi\rangle$. By contrast, vectorized operators differing by a global phase are not equivalent.   We further remark that when the operator is Hermitian, its evolution in Liouville space is equivalent to the quantum evolution of the corresponding physical observable in the Heisenberg picture.}

Let us now consider the Heisenberg evolution of an operator $O_0$ generated by a time-independent Hamiltonian $H$. For the moment, we do not restrict ourselves to observables, meaning that $O_0$ can also be non-Hermitian. In Liouville space, we can rewrite the Heisenberg equation as {$\partial_t \ket{O_t}=i\mathbb{L}\ket{O_t}$}, where the Liouvillian, which, with a slight abuse of notation, we shall continue to indicate as $\mathbb{L}$, takes the form
{
\begin{equation}\label{Liouvillian}
	\mathbb{L}=\frac{1}{\hbar}(H\otimes\mathbb{1}-\mathbb{1}\otimes H^T).
\end{equation}
}
A particularly useful object to quantify the displacement of the operator along the flow is the operator overlap
\begin{equation} \label{defoverlap}
\braket{O_0|O_t}=\frac{\Tr(O^\dagger_0O_t)}{\|O\|^2},
\end{equation}
which can be complex in general and becomes real when Hermitian operators are considered. A key observation is that the operator overlap is proportional to the infinite temperature autocorrelation function: we shall see how this feature will allow us to extend our result to finite-temperature autocorrelations functions as well. Note that $\braket{O_0|O_t}$ is closely related to the notion of operator fidelity introduced in the context of Loschmidt echoes and quantum phase transitions \cite{Wang09}. {Similarly to the fidelity of quantum states, the operator overlap \eqref{defoverlap} is potentially affected by the orthogonality catastrophe in many-body systems \cite{Bukov2019,Suzuki20,Delcampo21,Hamazaki2021,Hamazaki2022}, which would render it asymptotically small for large system sizes. Nevertheless, as shown in the random matrix example below, we can provide non-trivial bounds on the correlation functions.}

To measure how ``far'' the operator $O_t$ has flowed, we define the operator angle $\mathcal{L}_t$ between the vector $\ket{O_t}$ and the inital one $\ket{O_0}$:
\begin{equation}\label{defLt}
	\mathcal{L}_t\equiv\arccos{\operatorname{Re} \braket{O_0|O_t}}.
\end{equation}
This quantity defines a distance over the unitary flow $O_t=U_t^\dagger O_0 U_t$. Being the vectorized operator normalized as in Eq.~\eqref{vector}, the unitary flow in Liouville space lies on the unit sphere. Since $\operatorname{Re} \braket{O_0|O_t}$ is the Euclidean inner product, Eq.~\eqref{defLt} reduces to the angle between the corresponding real vectors in $\mathbb{R}^{2d^2}$ and is, therefore, a distance. We also note that $\mathcal{L}_t$ resembles the notion of Bures angle $\ell=\arccos{|\braket{\psi_0|\psi_t}|}$, but it does not reduce to it when the initial operator is chosen to be the projector over the pure state $\ket{\psi_0}$, $O_0=\ket{\psi_0}\bra{\psi_0}$. Moreover, let us stress that $\arccos{|\braket{O_0|O_t}|}$ would not be a good notion of distance between operators, as it would vanish every time they differ only by a phase. While physical states are defined up to an irrelevant global phase, the same is not true for operators, as global phases between observable are physical. Said differently, the fact that Liouville space is a Hilbert space but not a projective Hilbert space favors the use of the distance (\ref{defLt}) over the conventional Bures angle.

Now, let $\{\ket{i}\}$ be the energy eigenbasis, such that $H\ket{i}=E_i\ket{i}$, and let the initial operator be $O_0=\sum_{ij} O_{ij}\ket{i}\bra{j}$ in such basis. The Liouvillian is diagonal in the vectorized basis $\ket{i}\otimes\ket{j}$ and has the energy gaps $\Delta_{ij}\equiv E_i-E_j$ as diagonal entries. The operator overlap takes the form
\begin{equation} \label{overlap-total}
	\braket{O_0|O_t}=\frac{1}{\|O\|^2}\sum_{j,k}e^{i\Delta_{jk}t/\hbar}|O_{jk}|^2.
\end{equation}
Using trigonometric inequalities, we can derive two QSLs in terms of the operator overlap, which, as we will argue below, can be regarded as the generalization of the  Margolus-Levitin (ML) \cite{Margolus98} and Mandelstam-Tamm (MT) \cite{Mandelstam45} QSLs for Schr{\"o}dinger evolution. Indeed, by using that $\cos x \geq 1-\alpha|x|$, where the parameter $\alpha\approx 0.724$ is chosen such that $1-\alpha x$ is tangent to $\cos x$ for $x>0$ \cite{Andersson2019}, we can bound the real part of the overlap from below,
\begin{equation} \label{MLder}
\begin{split}
\operatorname{Re}\braket{O_0|O_t} &\geq 1- \frac{\alpha t}{\hbar\|O\|^2}\sum_{j,k}|\Delta_{jk}||O_{jk}|^2\\
&=1-\alpha\braket{|\mathbb{L}|}t,
\end{split}
\end{equation}
where, {for an arbitrary operator $A$, we define $|A|=\sqrt{A^\dagger A}$} and the brackets in the right-hand side stand for the expectation value over the vectorized operator: $\braket{\cdot}=\braket{O_0|\cdot|O_0}$. Further, its time derivative $\operatorname{Re} \braket{O_0|\dot{O}_t}$ can be upper bounded by making use of the inequality $-x^2\leq x\sin x\leq x^2$, valid $\forall x$:
\begin{equation}\label{O'upperbound2}
	 |\operatorname{Re} \braket{O_0|\dot{O}_t}|\leq\frac{t}{\hbar^2\|O\|^2}\sum_{j,k}\Delta_{jk}^2|O_{jk}|^2= \braket{\mathbb{L}^2}t.
\end{equation}
The derivative of the operator overlap is related to that of the operator angle $\mathcal{L}_t$ by
\begin{equation} \label{diff-approach}
\operatorname{Re}\braket{O_0|\dot{O}_t}=-\sin{(\mathcal{L}_t)}\, \dot{\mathcal{L}_t},
\end{equation}
which, after time integration, when combined with Eq.~\eqref{O'upperbound2}, yields
\begin{equation} \label{MTder}
1-\cos{\mathcal{L}_t} \leq \frac{ \braket{\mathbb{L}^2}}{2}t^2.
\end{equation}
Therefore, from Eqs.~\eqref{MLder} and \eqref{MTder} we obtain the two following QSLs, formulated in terms either of the operator overlap or the operator angle:
\begin{align} \label{QSL-ML}
t &\geq \frac{1-\operatorname{Re}\braket{O_0|O_t}}{\alpha\braket{|\mathbb{L}|}}=\frac{1-\cos{\mathcal{L}_t}}{\alpha\braket{|\mathbb{L}|}}, \\
\label{QSL-MT}
t &\geq  \sqrt{ \frac{ 2(1-\operatorname{Re}\braket{O_0|O_t} )}{ \braket{\mathbb{L}^2}} }=\sqrt{ \frac{2(1- \cos{\mathcal{L}_t)}} { \braket{\mathbb{L}^2}} }.
\end{align}
These results identify $\braket{|\mathbb{L}|}$ and $\sqrt{ \braket{\mathbb{L}^2} }$ as upper bounds on the speeds of the operator flow. We stress that the quantities $\braket{|\mathbb{L}|}$ and $\braket{\mathbb{L}^2}$ are time-independent under unitary dynamics when the evolution is generated by a time-independent Hamiltonian. Moreover, we note that if $O_0$ is Hermitian, the operator overlap is real at any time and $\braket{\mathbb{L}}=0$ by parity so that $\braket{\mathbb{L}^2}=(\Delta \mathbb{L})^2$ is the variance of the Liouvillian.
The analogy with ML and MT bounds, at least for flows of observables, is already evident. Both for state and operator evolution, the relevant time scale is given in terms of the mean and the variance of the generator of evolution, for ML and MT bounds, respectively. In the case of states, the generator is the Hamiltonian $H$, while in Liouville space the dynamics of operator flows is generated by the Liouvillian $\mathbb{L}$. We observe that, in the case of operator evolution, the ML bound \eqref{QSL-ML} is given in terms of $\braket{|\mathbb{L}|}$ rather than the mean of the generator $\braket{\mathbb{L}}$, as this one vanishes for Hermitian operators. Let us further note that, by using other trigonometric inequalities, one can derive analogous QSLs that are proportional to Eqs.~\eqref{QSL-ML}-\eqref{QSL-MT} through a numerical constant smaller than one, thus yielding a weaker result, see App.~\ref{AppQSLTrig}.

It is instructive to identify which operators maximize the upper bounds $\braket{|\mathbb{L}|}$ and $\sqrt{ \braket{\mathbb{L}^2} }$ on the speed of the flow for a given Hamiltonian, as these will undergo the fastest operator growth. By looking at Eqs.~\eqref{O'upperbound2} and \eqref{MLder}, and noting that $(|O_{ij}|/\|O\|)^2\leq1$ defines a proper probability distribution over the energy states pairs, it is clear that the operators $O_{\text{max}}$ flowing at the maximal speed are the ones whose non-zero elements are only between energy eigenstates with the maximum gap $|\Delta_{\text{max}}|=E_{\text{max}}-E_0$, where $E_{\text{max}}$ and $E_0$ are the highest and the lowest energy eigenvalues, respectively. This maximal speed for operators is the analog  of the one identified by the dual ML bound, recently introduced for state evolution \cite{Ness2022}. If these levels are non-degenerate, the fastest operator will be of the form $O_{\text{max}}=\mu\ket{E_{\text{max}}}\bra{E_0}+\nu\ket{E_0}\bra{E_{\text{max}}}$ for some complex constant $\mu$ and $\nu$ ($\mu=\nu^*$ for observables). This is analogous to the well-known result for states \cite{Levitin2009}.

Remarkably, the above QSLs identify a universal crossover between two different time regimes. At early times $t\leq \tau_\textrm{c}$, being $\tau_\textrm{c}=2 \alpha \braket{|\mathbb{L}|}/\braket{\mathbb{L}^2}$ the crossover time, the decay of the operator overlap is governed by a quadratic MT bound
\begin{equation}
\operatorname{Re}\braket{O_0|O_t} \geq 1- \frac{\braket{\mathbb{L}^2}}{2} t^2,
\end{equation}
while for times $t\geq \tau_\textrm{c}$ the linear ML bound 
\begin{equation}
\operatorname{Re}\braket{O_0|O_t} \geq 1- \alpha\braket{|\mathbb{L}|} t
\end{equation}
becomes tighter. A similar crossover was experimentally observed for the state fidelity of a single atom in an optical trap \cite{Ness2021}.

Furthermore, let us note that the QSLs \eqref{QSL-ML} and \eqref{QSL-MT} can be recast in terms of the Hamiltonian $H$, which generates the time evolution in the Hilbert space, rather than the Liouvillian, i.e.,~the generator of evolution in Liouville space. This reformulation will prove advantageous in expressing the relevant timescales as thermal expectation values. Indeed, $\sqrt{\braket{\mathbb{L}^2}}$ is proportional to the norm of the operator velocity
\begin{equation} \label{DeltaL-velocity}
\sqrt{\braket{\mathbb{L}^2}}=\frac{1}{\hbar}\frac{\|[H,O_t]\|}{\|O\|}=\frac{\|\partial_tO\|}{\|O\|},
\end{equation}
which, as emphasized, is time-independent under unitary dynamics. Regarding the ML QSL, by using that $|\Delta_{jk}| \leq E_j+ E_k-2E_0$ in terms of  the ground state energy $E_0$, we find
\begin{equation}
\braket{|\mathbb{L}|} \leq \frac{1}{\hbar}\frac{\Tr (O_0^\dagger \{H-E_0, O_0\} )}{\|O\|^2}.
\end{equation}
Therefore, the ML QSL can be recast as
\begin{equation} \label{QSL-MLtr} 
t \geq \hbar\frac{ \|O\|^2 }{\alpha} \frac{1-\operatorname{Re}\braket{O_0|O_t}}{\Tr (O_0^\dagger \{H-E_0, O_0\}  )},
\end{equation}
which is generally weaker than the original Liouvillian bound \eqref{QSL-ML}. We shall make use of these results in the next section, to bound the rate of change of autocorrelation functions.

Finally, if we choose $O_0=\ket{\psi_0}\bra{\psi_0}$ and let it evolve backward in time, thus recovering the corresponding forward time evolution of the state $\ket{\psi_0}$ in the Schr{\"o}dinger picture, we find that the bounds \eqref{QSL-ML} and \eqref{QSL-MT} become proportional to the standard ML and MT QSLs in the case of orthogonal state evolution, thus justifying the given interpretation, see App.~\ref{AppQSLSchro}. We note that the proportionality constant is smaller than one, meaning that our bounds are not violated. However, they cannot be tight for state evolution. Furthermore, the result \eqref{QSL-ML-states} can be extended to driven dynamics, under the assumption that the energy eigenvectors are stationary, see App.~\ref{MLOpFlowDriven}.

The general results derived in this section can be applied in a variety of theoretical settings. In particular, as shown below, the QSLs on operator flows bound the decay of autocorrelation functions, thus providing constraints to the dynamical susceptibility in linear response theory and the quantum Fisher information in quantum metrology.

\section{QSL on autocorrelation functions} \label{sec:autocorr}

Solving the dynamics of an arbitrary  many-body quantum system is generally a demanding task. In many-body physics, the central objects that characterize the dynamics, determining for example the linear response \cite{Kubo1957}, are the two-point time-correlation functions, whose explicit form is generally unknown. In particular, the so-called autocorrelation function
\begin{equation} \label{autocorr}
C_O(t)=\Tr(O^\dagger_tO_0\rho)
\end{equation}
determines the operator growth of $O_0$ in Krylov space and therefore accounts for the build-up of the corresponding notion of operator complexity \cite{Parker2019,Barbon2019,Rabinovici2021,Caputa2021,Dymarsky2021}.

Our QSLs on operators provide easily computable lower bounds to these quantities.  In the following, we take $O_0$ to be Hermitian, $\rho$ to be a stationary state, $[\rho,H]=0$, and the Hamiltonian $H$ to be time independent. In practical applications, $\rho$ is often chosen to be the Gibbs state $e^{-\beta H}/Z$ at inverse temperature $\beta$, with $Z=\Tr e^{-\beta H}$, but this assumption is not necessary for the derivation of the results below. Now, let us note that this autocorrelation function can be rewritten as the Hilbert-Schmidt inner product
\begin{equation} \label{autocorr-HS}
C_O(t)=\braket{\tilde{O}_t,\tilde{O}_0}
\end{equation}
between a non-Hermitian operator $\tilde{O}_0$ and its time-evolved operator $\tilde{O}_t$, defined as follows:
\begin{equation}
\tilde{O}_0\equiv O_0\sqrt{\rho}, \quad \tilde{O}_t = U^\dagger\tilde{O}_0U=O_t\sqrt{\rho}.
\end{equation}
By making the commutator explicit and using that $[\rho,H]=0$, we obtain $\|[H,\tilde{O}_0]\|^2=\hbar^2  { \braket{\dot{O}^2_t}_0}$, where the brackets {$\braket{\cdot}_0$} stand for the expectation value with respect to $\rho$. We note that the characteristic velocity {$\braket{\dot{O}^2_t}_0$} of the operator flow is time independent and thus does \textit{not} require solving for $O_t$. Being $\|\tilde{O}_0\|^2 = C_O(0)$ the norm of $\tilde{O}_t$, the autocorrelation function becomes $C_O(t)=C_O(0)\braket{\tilde{O}_t|\tilde{O}_0}$. Substituting this expression into Eqs.~\eqref{QSL-MT} and \eqref{DeltaL-velocity}, we can recast our MT QSL as a lower bound on the symmetrized autocorrelation function
\begin{equation} \label{Autocorr-MT}
\operatorname{Re} C_O(t) = \frac{1}{2} \braket{ \{O_t,O_0 \} }_0 \geq C_O(0) - \frac{1}{2} {\braket{\dot{O}^2_t}_0} t^2,
\end{equation}
which, as also occurs for the decay of the state fidelity, corresponds to a short-time Taylor expansion up to the second order. This early timescale was also identified in \cite{Alhambra2020} and found to be proportional to the equilibration timescale at late times. While the MT QSL accounts for the short-time quadratic decay of the autocorrelation function, the ML QSL \eqref{QSL-MLtr} yields a linear decay instead,
\begin{equation} \label{Autocorr-ML}
\operatorname{Re} C_O(t) \geq C_O(0) - \frac{\alpha}{\hbar} { \braket{O_0\{H-E_0,O_0\}}_0 }t,
\end{equation}
which identifies a new timescale. Again, as for the evolution of the operator overlap, we observe a crossover between the MT and the ML regimes, occurring at the time
\begin{equation} \label{Autocorr-Im}
\tau_\textrm{c}=\frac{2\alpha}{\hbar} {\frac{\braket{O_0\{H-E_0,O_0\}}_0}{ \braket{\dot{O}^2_t}_0}},
\end{equation}
which is illustrated explicitly in Figs.~\ref{fig-2-level} and \ref{fig-RMT-Op} for a two-level system and a random matrix Hamiltonian, respectively.

Furthermore, as the operator dynamics is governed by the full autocorrelation function \eqref{autocorr}, one wishes to have a  bound also on the anti-symmetrized, imaginary autocorrelation function $\braket{[O_t,O_0]}_0$. This quantity determines the linear response of the operator when the Hamiltonian is perturbed with an external, time-dependent driving \cite{Kubo1957}. Let us then discuss an ML-type upper bound on the imaginary part of the autocorrelation function. This can be achieved by noting that the imaginary part of the operator overlap \eqref{overlap-total} is upper bounded by the averaged Liouvillian $\braket{|\mathbb{L}|}$,
{
\begin{equation} 
|\operatorname{Im} \braket{O_0|O_t}|\leq \braket{|\mathbb{L}|}t,
\end{equation}
}
as one can see by using that $-|x|\leq \sin{x} \leq |x| $ $\forall x$. Now, being $\operatorname{Im} C_O(t)=-C_O(0)\operatorname{Im}\braket{\tilde{O}_0|\tilde{O}_t}$, we derive
\begin{equation} \label{bound-ImCorr}
|\operatorname{Im}C_O(t)|  \leq {\braket{O_0 \{ H-E_0,O_0 \} }_0} \frac{t}{\hbar}.
\end{equation}
This result will be used below to derive analogous bounds on the linear response under an external perturbation and the thermal quantum Fisher information associated with an arbitrary observable $O$.

\subsection{The two-level system}

Before considering further applications of our results, let us analyze a simple model for which we can compute both the bounds \eqref{Autocorr-MT}-\eqref{Autocorr-Im} and the actual autocorrelation function $C_O(t)$, thus illustrating explicitly the existence of the aforementioned crossover.

\begin{figure}[t]
	\centering
	
	\includegraphics[width=0.4\textwidth]{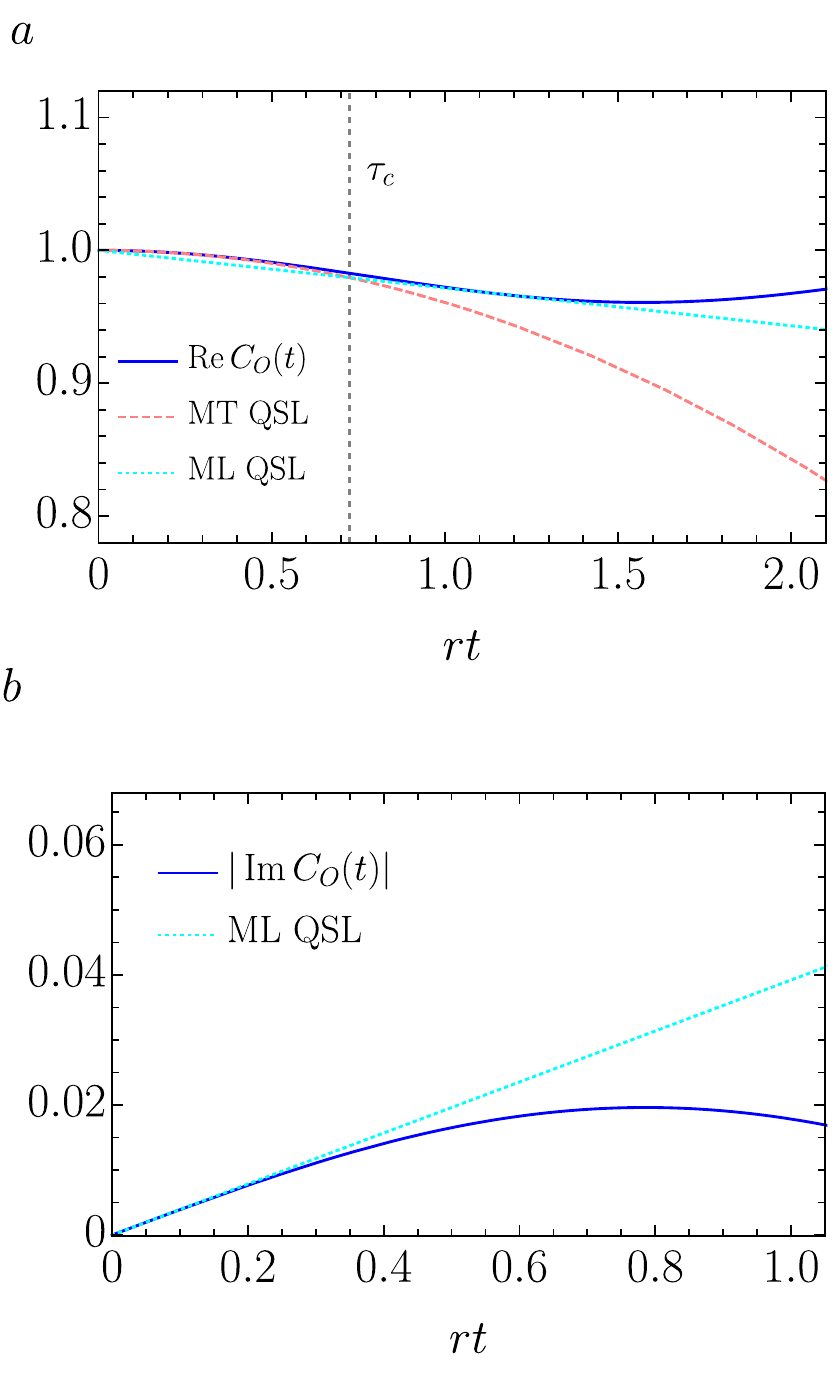}

	\caption{Comparison of the real (a) and imaginary (b) parts of the correlation function with the MT and ML QSLs given by Eqs.~\eqref{Autocorr-MT}, \eqref{Autocorr-ML} and \eqref{bound-ImCorr} for a two-level system. The parameters are chosen as $a=10$, $c=b=1$, and the inverse temperature is $\beta=10$. The identity in the Hamiltonian \eqref{H-2lev} plays no role so that we can fix $k=0$ without loss of generality. (a) The initial decay of the symmetric correlation function undergoes a crossover from a regime dominated by the MT QSL to a regime in which the ML QSL becomes tighter. The vertical line corresponds to the crossover time $\tau_\textrm{c}$. (b) The onset of the antisymmetric contribution is characterized by the ML QSL at short times.}
	\label{fig-2-level}
	
\end{figure}

Let us consider a two-level Hamiltonian
\begin{equation} \label{H-2lev}
H= k\mathbb{1} + \overrightarrow{r} \cdot \overrightarrow{\sigma},
\end{equation}
where $\overrightarrow{\sigma}=(\sigma_x,\sigma_y,\sigma_z)$ is the Pauli matrix vector and $\overrightarrow{r}=(a,b,c)$. Let us choose $O_0=\sigma_x$ as the initial operator and $\rho=e^{-\beta H}/Z$ as the state of the system. Then, using natural units $\hbar=1$, the symmetric autocorrelation function is given by
 \begin{equation} \label{Recorr}
\operatorname{Re} C_O(t)=\frac{a^2+(b^2+c^2)\cos{2rt}}{r^2},
\end{equation}
where $r\equiv |\overrightarrow{r}|=\sqrt{a^2+b^2+c^2}$, the temperature dependence being contained only in the antisymmetric part
\begin{equation} \label{Imcorr}
\operatorname{Im} C_O(t)=\frac{b^2+c^2}{r^2}\tanh{\beta r}\sin{2rt}.
\end{equation}
Moreover, the MT and ML time scales have the following expressions respectively,
\begin{equation}
{\braket{\dot{O}^2_t}_0} = 4(b^2+c^2)
\end{equation}
and
\begin{equation}
{\braket{O_0\{H-E_0,O_0\}}_0} = \frac{2}{r} \left( \frac{2a^2}{1+e^{2\beta r}} + b^2+c^2 \right).
\end{equation}
These can be inserted into Eqs.~\eqref{Autocorr-MT} and \eqref{Autocorr-ML} to compute lower bounds on $\operatorname{Re} C_O(t)$ and $\operatorname{Im} C_O(t)$, as shown in Fig.~\ref{fig-2-level}. The symmetric, real part is the only nonvanishing term at the initial time and undergoes a decay in parallel with the onset of the antisymmetric, imaginary contribution. This decay is initially captured by the MT QSL, as expected from Taylor expansion, while after the crossover time $\tau_\textrm{c}$, the ML QSL becomes tighter.

\begin{figure}[t] 
	\centering	
	\includegraphics[width=0.4\textwidth]{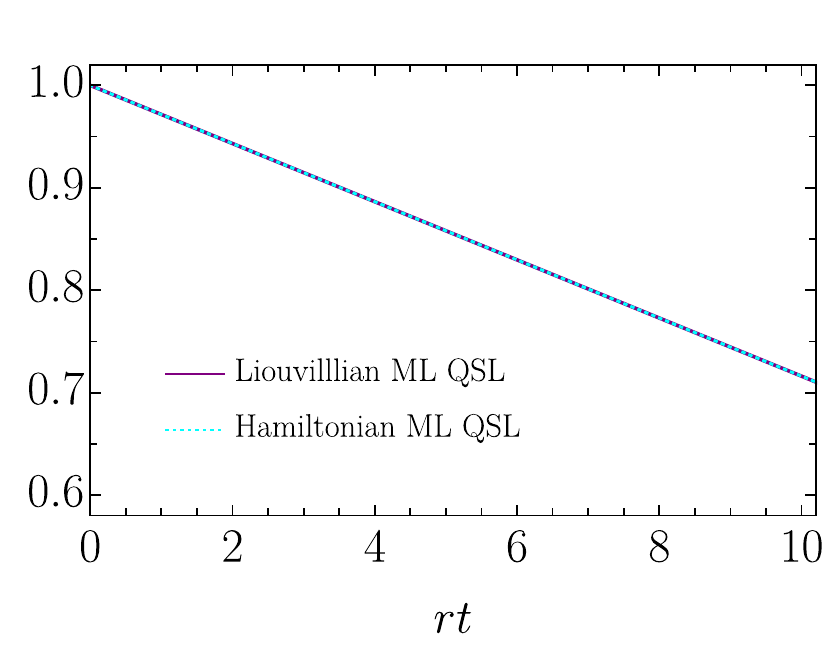}
	\caption{The Liouvillian \eqref{QSL-ML} and Hamiltonian \eqref{QSL-MLtr} formulations of the ML bound for a two-level system are shown to be equivalent in the time window of interest. The parameters are chosen as $a=10$, $c=b=1$ and the inverse temperature is $\beta=10$.}
	\label{fig-MLcomparison}
\end{figure}

Finally, let us note that, for the parameters chosen ($a=10$, $c=b=1$ and $\beta=10$), the two ML QSLs, formulated in terms either of the Hamiltonian \eqref{QSL-MLtr} or the Liouvillian \eqref{QSL-ML}, are equivalent, as shown in Fig.~\ref{fig-MLcomparison}. Indeed, in the latter case, the velocity is given by
\begin{equation}
C_O(0)\braket{\tilde{O}| \, |\mathbb{L}| \, |\tilde{O} } = \frac{2}{r}( b^2+c^2),
\end{equation}
so that the difference between the two ML velocities is
\begin{equation}
\frac{4a^2}{r}\frac{\alpha}{1+e^{2\beta r}} \ll 1.
\end{equation}
However, if one increases the temperature sufficiently the Liouvillian bound  gives a tighter result.

\subsection{Random matrix example}

\begin{figure}[t]
	\centering
	
	\includegraphics[width=0.4\textwidth]{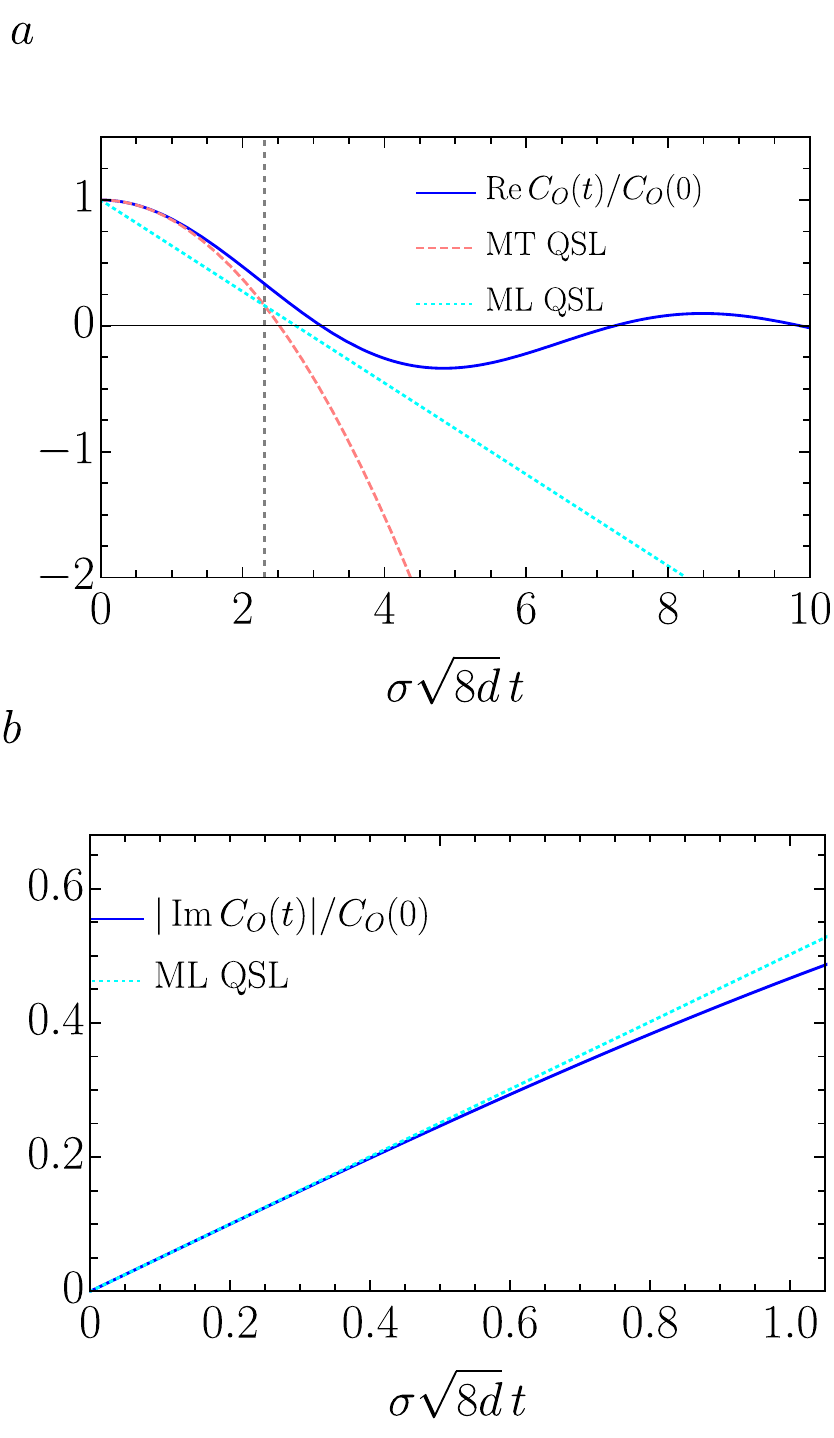}

	\caption{Comparison of the real (a) and imaginary (b) parts of the (normalized) correlation function with the MT and ML QSLs given by Eqs.~\eqref{Autocorr-MT}, \eqref{Autocorr-ML} and \eqref{bound-ImCorr}, for $H,\,O_0\in$ GOE with standard deviation $\sigma=1$ and dimension $d=200$. The QSLs have been  normalized by dividing by $C_0(0)$. (a) The initial decay of the symmetric correlation function undergoes a crossover from a regime dominated by the MT QSL to a regime in which the ML QSL becomes tighter. The vertical line corresponds to the crossover time $\tau_\textrm{c}$. (b) The onset of the antisymmetric contribution is characterized by the ML QSL at short times.}
	\label{fig-RMT-Op}
	
\end{figure}

As already stressed, the above scenario consisting of an early quadratic decay governed by MT \eqref{Autocorr-MT}, a crossover, and a subsequent time window in which ML \eqref{Autocorr-ML} becomes tighter, proves to be a general feature of autocorrelations in isolated quantum systems.
{We further illustrate this scenario in the following generic setting. First, we sample the Hamiltonian $H$ from the Gaussian Orthogonal Ensemble (GOE), with standard deviation $\sigma=1$ and dimension $d=200$; we subsequently diagonalize $H$ and thus fix as a reference basis the energy eigenbasis. We then construct the initial operator, using this basis, by sampling it from the same GOE as the Hamiltonian. Alternatively, we could sample the operator and the Hamiltonian using the same computational basis.} The comparison between the autocorrelation function and the speed limits is shown in Fig.~\ref{fig-RMT-Op}.

While the tangent character of the ML QSL to the real curve after the crossover is a feature specific to the qubit case (see Fig.~\ref{fig-2-level}), we observe that the divergence of the bound does not keep increasing when considering an increasing Hilbert space dimension. ML-type bounds were recently found to lose tightness for higher dimensions than a qubit also in \cite{Ness2022}.

\section{Dynamical susceptibilities}

Thermal correlation functions  determine the non-equilibrium response of an observable at the first order in the perturbation \cite{Kubo1957,Tuckerman}. Therefore, the results illustrated in the previous section allow us to bound the linear response and, in particular, the dynamical susceptibility, which is the quantity that characterizes the response of a system to an external perturbation.

Let us then consider the situation in which a time-independent Hamiltonian $H_0$ is perturbed by an external, time-dependent driving,
\begin{equation}
H(t) = H_0+ \lambda V f(t),
\end{equation}
where the perturbation operator $V$ does not depend explicitly on time, $\lambda$ is a real positive constant that quantifies the strength of the perturbation and the driving force $f(t)$, which can be taken to be $|f|\leq1$, is assumed to vanish for $t\leq 0$. Let the initial state of the system be the thermal Gibbs state $\rho_0$, relative to the unperturbed Hamiltonian $H_0$, at inverse temperature $\beta$. We are interested in determining the linear response of an observable $A$, namely the shift of its expectation value at time $t$ from the initial equilibrium value, $\braket{A}_t-\braket{A}_0$, where 
\begin{equation}
\braket{A}_0=\Tr{\left(A\rho_0\right)}, \quad \braket{A}_t=\Tr{\left[\rho_I(t)A_I(t)\right]}.
\end{equation}
Here, the operators are evaluated in the interaction picture
\begin{equation} \label{int-op}
A_I(t)=U^\dagger_0(t) A U_0(t), \quad \rho_I(t)=U_V(t) \rho_0 U_V^\dagger(t),
\end{equation}
with
\begin{equation}
U_0(t)=e^{-iH_0 \frac{t}{\hbar} }, \quad U_V(t)=\mathcal{T} e^{-i\frac{\lambda}{\hbar}\int_{0}^{t} V_I(s)f(s) ds}.
\end{equation}
To the first order in $\lambda$, the linear response is given by the celebrated Kubo formula \cite{Kubo1957}
\begin{equation} \label{Kubo}
\braket{A}_t-\braket{A}_0 \simeq \frac{\lambda}{i\hbar}\int_{0}^{t} \braket{[A_I(t-s),V_I(0)]}_0 f(s) ds.
\end{equation} 
Thus, the non-equilibrium response, at the linear order, is determined by an equilibrium correlation function. As already stressed above, computing correlation functions at all times is generally a challenging task, equivalent to solving the dynamics. Therefore, having universal, model-independent bounds on these quantities is extremely useful. We next present several ways in which one can bound the right-hand side of Eq.~\eqref{Kubo}.

The linear response \eqref{Kubo} can be rewritten in terms of the so-called dynamical susceptibility
\begin{equation} \label{susc}
\chi_{AV}(t)=-i\frac{1}{\hbar} \theta(t) \braket{[A_I(t),V_I(0)]}_0,
\end{equation}
where $\theta(t)$ is the Heaviside function, as
\begin{equation} \label{Kubo2}
\braket{A}_t- \braket{A}_0 = \lambda \int_{-\infty}^{\infty} \chi_{AV}(t-s) f(s) ds.
\end{equation} 
The dynamical susceptibility $\chi_{AV}(t)$, which vanishes for a negative argument (i.e., before the external perturbation is applied), expresses the causal linear response of the system and is a real quantity. Its absolute value can be upper bounded using the Heisenberg uncertainty relation $|\braket{ [A,V] }| \leq 2\Delta A \Delta V$, which yields the constant bound
\begin{equation} \label{bound-susc1}
|\chi_{AV}(t)|\leq \frac{2}{\hbar} \theta(t) \Delta_0A \Delta_0V,
\end{equation}
where $\Delta^2_0 A$ is the variance with respect to $\rho_0$. As the latter is time-independent in the interaction picture,  we drop the subscript $I$.
Alternatively, one can apply the Bogoliubov inequality \cite{Ueda}
\begin{equation} \label{Bog}
|\braket{[A,V]}_0|^2 \leq \frac{\braket{A^2}_0\braket{[V,[H_0,V]]}_0}{k_BT}.
\end{equation}
This yields  a different upper bound with an explicit dependence on the temperature $T$
\begin{equation} \label{bound-susc-Bog}
|\chi_{AV}(t)|\leq  \frac{2\theta(t)}{\hbar} \sqrt{\frac{T}{T_B}} \Delta_0 V \Delta_0 A,
\end{equation}
where we have defined the characteristic temperature
\begin{equation} \label{TBog}
T_B \equiv \frac{ \braket{A^2}_0\braket{[V,[H_0,V]]}_0 }{4 k_B (\Delta_0 A \Delta_0 V)^2}.
\end{equation}
{Both (\ref{bound-susc1}) and (\ref{bound-susc-Bog}) upper bound the modulus of the dynamical susceptibility in terms of the equilibrium fluctuations of the perturbation operator $V$ and the observable $A$ in which the response is studied.}
We see that at low temperature $T\leq T_B$ the Bogoliubov bound \eqref{bound-susc-Bog} is tighter than the Heisenberg bound \eqref{bound-susc1}, while the contrary holds at higher temperatures.

In certain experimental settings, one may be interested in quantifying the response of the perturbation operator $V$ itself. For example, this is the case of the magnetic susceptibility in magnetic resonance experiments, where both $V$ and the observable of interest $A$ are given by the magnetization \cite{Mazenko}. In such case, being $A=V$, the dynamical susceptibility $\chi_{VV}(t)$ becomes proportional to the anti-symmetrized autocorrelation function $\braket{[V_I(t),V_I(0)]}_0=2i\operatorname{Im} C_V(t)$, and therefore one can make use of the techniques that we have illustrated above to bound autocorrelations functions utilizing QSLs for operators. The ML QSL \eqref{Autocorr-Im} yields the following upper bound on the dynamical susceptibility
\begin{equation} \label{bound-susc2}
|\chi_{VV}(t)|\leq\frac{t}{\tau_{QSL}^3} 2\theta(t)\hbar,
\end{equation}
where we have introduced a new time scale $\tau_{QSL}=\hbar \braket{V \{ H_0-E_0,V \} }^{-1/3}_0$, {$E_0$ being the ground state energy of the unperturbed Hamiltonian $H_0$}. Unlike the previous bounds \eqref{bound-susc1} and \eqref{bound-susc-Bog}, Eq.~\eqref{bound-susc2} contains an explicit linear dependence on time. This implies that the QSL approach is the most efficient at early enough times. However, as we let the system evolve in time, the QSL bound \eqref{bound-susc2} no longer governs the dynamics, and one needs to consider different approaches. In particular, at lower temperature $T \leq T_B$, the QSL bound \eqref{bound-susc2} ceases to be the tightest one at the time
\begin{equation}
\tau_B= \Big(\frac{\Delta_0V}{\hbar}\Big)^2 \sqrt{\frac{T}{T_B}} \tau^3_{QSL}
\end{equation}
and for $t\geq \tau_B$ the Bogoliubov bound \eqref{bound-susc-Bog} gives the better description. By contrast, at high temperatures $T\geq T_B$, the crossover is between the QSL bound and the Heisenberg bound \eqref{bound-susc1}, and it occurs at the time
\begin{equation}
\tau_H= \Big(\frac{\Delta_0V}{\hbar}\Big)^2 \tau^3_{QSL}.
\end{equation}
The bounds \eqref{bound-susc1},\eqref{bound-susc-Bog} and \eqref{bound-susc2} find applications in many experimentally relevant settings \cite{Mazenko}, of which we give two concrete examples below.

\subsection{Examples}

The situation in which a system is subject to an external perturbation is widespread in physics. A paradigmatic example is that of a system composed of $N$ charged particles and perturbed with a uniform, time-dependent electric field $\overrightarrow{E}(t)$ \cite{Mazenko}. In this case, the preferred observable to characterize the response of the system is the current flow $\overrightarrow{J}$ 
and the corresponding dynamical susceptibility is the electrical conductivity. The perturbed Hamiltonian takes the form
\begin{equation} 
H(t)=H_0-\overrightarrow{R}\cdot \overrightarrow{E}(t),
\end{equation}
where $\overrightarrow{R}$ is the electric dipole moment in the origin
\begin{equation}
\overrightarrow{R}\equiv \sum_{n}^{N} q_n \overrightarrow{r}_n,
\end{equation}
with $q_n$ and $\overrightarrow{r}_n$ being the charge and the position of the $n$-th particle. The net current vanishes at equilibrium, $\braket{\overrightarrow{J}}_0=0$, and the perturbed expectation value of the $i$-th spatial component $J_i$ at time $t\geq0$ is expressed, at the linear order, as \cite{Mazenko}
\begin{equation}
\braket{J_i}_t=\sum_j \int_{-\infty}^{\infty} \sigma_{ij}(t-s) E_j(s) \, ds,
\end{equation}
where $\sigma_{ij}\equiv \chi_{J_i R_j}$ is the electrical conductivity tensor. By using the bounds \eqref{bound-susc1} and \eqref{bound-susc-Bog} we are able to derive the following constraints on $\sigma_{ij}$:
\begin{align}
 |\sigma_{ij}(t)| &\leq \frac{2}{\hbar} \theta(t) \Delta_0J_i \Delta_0 R_j,\\
 |\sigma_{ij}(t)| &\leq \frac{\theta(t)}{\hbar} \sqrt{\frac{\braket{J_i^2}_0\braket{[R_j,[H_0,R_j]]}_0}{k_BT} }, \label{condBog}
\end{align}
{where, in the first equation, we note that that the net current vanishes at equilibrium, i.e., $\Delta_0J_i=\sqrt{\braket{J_i^2}_0}$.}
{Thus, the application of the Heisenberg bound \eqref{bound-susc1} yields an experimentally testable upper bound to the modulus of each component of the electrical conductivity tensor $|\sigma_{ij}(t)|$ in terms of the equilibrium fluctuations of the current flow $\overrightarrow{J}$ along $i$-th axis  and  the $j$-th component of the electric operator $\overrightarrow{R}$ to which the applied external field $\overrightarrow{E}(t)$ is coupled.}
{Provided that the unperturbed Hamiltonian $H_0$ depends on the momenta $\overrightarrow{p}_n$ through the usual kinetic term $\sum_n \frac{\overrightarrow{p}_n^2}{2m}$, where we have taken the particles to have equal masses $m$, we can evaluate the double commutator appearing in the Bogoliubov bound \eqref{condBog}. Thus, by using the canonical commutation relations, we obtain
\begin{equation}
|\sigma_{ij}(t)| \leq
\theta(t) \Delta_0J_i \sqrt{\frac{\sum_n q_n^2}{m k_B T}},
\end{equation}
where we note that temperature dependence is also contained in the current fluctuations $\Delta_0J_i$. 
}

Another experimental application in which the theory of linear response provides a useful approach is given by magnetic resonance experiments \cite{Pake}. In this case, the central quantity is the magnetic susceptibility. 
Consider a paramagnetic system, initially aligned along a constant magnetic field $\overrightarrow{B}$ and subsequently perturbed with a weak time-dependent field $\overrightarrow{h}(t)$ so that the total Hamiltonian reads
\begin{equation}
H(t)=-\overrightarrow{M}\cdot(\overrightarrow{B}+ \overrightarrow{h}(t)).
\end{equation}
The magnetization of the system is perturbed from its initial equilibrium value $\braket{\overrightarrow{M}}_0=\chi_0\overrightarrow{B} $, where $\chi_0$ is the static susceptibility, and the linear response of its $i$-th component can be expressed using the Kubo formula
\begin{equation}
\braket{M_i}_t - \braket{M_i}_0=\sum_j \int_{-\infty}^{\infty} \chi_{M,ij}(t-s) h_j(s) \,ds,
\end{equation}
where $\chi_{M,ij}(t)\equiv\chi_{M_i M_j}(t)$ is the magnetic susceptibility tensor. The bounds \eqref{bound-susc1} and \eqref{bound-susc-Bog} now yield
\begin{align}
|\chi_{M,ij}(t)| &\leq \frac{2}{\hbar} \theta(t) \Delta_0M_i \Delta_0M_j,\\
|\chi_{M,ij}(t)| &\leq \frac{\theta(t) }{\hbar} \sqrt{\frac{\braket{M_i^2}_0\braket{[M_j,[H_0,M_j]]}_0}{k_BT}} . \label{magnBog}
\end{align}
{As in the previous example, the absolute value of each component of the magnetic susceptibility $|\chi_{M,ij}(t)|$ is upper bounded in terms of the equilibrium fluctuations of the response $M_i$ and the perturbation operator, which in this case is given by the magnetization itself, along the $j$-th axis. Let us consider as a simple example the case of $N$ decoupled spins, subject to the external magnetic field $\overrightarrow{B}$. Then $M_i=\gamma\sum_n \sigma^{(n)}_i$, where the proportionality constant $\gamma$ has the dimension of a magnetic dipole moment, and the unperturbed ground state energy is $E_0=-N\gamma|\overrightarrow{B}|$. By using the commutation relations for the Pauli matrices $\sigma_i$, we evaluate the Bogoliubov bound \eqref{magnBog} on the dynamical response as
\begin{equation}
|\chi_{M,ij}(t)| \leq \frac{2\gamma\theta(t) }{\hbar} \sqrt{\frac{\chi_0 \braket{M_i^2}_0 (|\overrightarrow{B}|^2-B_j^2) }{k_BT}}.
\end{equation}
We note that this bound is consistent with the fact that $\chi_{M,ij}$ vanishes whenever the static magnetic field is along the $j$-th axis, which follows from the vanishing of the averages $\braket{M_i}_0$ along directions orthogonal to $\overrightarrow{B}$.}
Moreover, the diagonal magnetic susceptibilities can be upper bounded through the QSL approach \eqref{bound-susc2} to find
\begin{equation} 
|\chi_{M,ii}(t)| \leq \frac{2}{\hbar^2} \theta(t) \braket{M_i \{ H_0-E_0, M_i \} }_0 t.
\end{equation}

\section{Bounds on the Quantum Fisher Information} \label{sec:Fish} 
We next turn our attention to the application of QSL on operator flows in quantum metrology. In this context, 
the quantum Fisher information $F_Q$ associated with a Hermitian operator $O$ quantifies the maximal precision with which we can estimate the phase $\theta$ that parameterizes the \textit{unitary flow}, of a given quantum state $\rho$, generated by the operator $O$. In other words, $F_Q$ measures the distinguishability of the ``initial'' state $\rho_0$ from the one transformed by the unitary flow
\begin{equation}
\rho_\theta= e^{-i\theta O} \rho_0 e^{i\theta O}.
\end{equation}
Let us note the change in the perspective: instead of looking at the \textit{time} unitary flow that the observable $O$ of interest undergoes under the action of the Hamiltonian that generates the dynamics, we are now considering the unitary flow (in a different parameter $\theta$) of a given quantum state $\rho$ under the action of $O$, that generates the state transformation. Remarkably, these two approaches are closely related, as the dynamical susceptibility $\chi_{OO}$ obtained using the first framework can be related to the quantum Fisher information $F_Q$ \cite{Hauke2016,Brenes20}.

More precisely, for a thermal state $\rho$ at temperature $T$, the following result on the quantum Fisher information has been shown by Hauke \textit{et.~al} in \cite{Hauke2016}
\begin{equation} \label{Hauke}
F_Q(T)= -\frac{4}{\pi} \int_{0}^\infty d\omega \tanh\big( \frac{\hbar\omega}{2k_BT} \big) \operatorname{Im} \tilde{\chi}_{OO} (\omega,T),
\end{equation}
where $\tilde{\chi}_{OO} (\omega,T)$ is the Fourier-transformed dynamical susceptibility, defined as
\begin{equation}
\tilde{\chi}_{OO} (\omega,T)=  \int_{0}^\infty e^{i\omega t} \chi_{OO}(t,T)\,dt,
\end{equation}
with $\chi_{OO}(t,T)$ being defined in Eq.~\eqref{susc} for $A=V=O$.

Now, the upper bound \eqref{bound-ImCorr} we have derived above on the anti-symmetrized autocorrelation function $\operatorname{Im} C_O$, together with the result \eqref{Hauke}, provides an upper bound on the quantum Fisher information $F_Q$. To this end, let us reverse the order of the integrals in Eq.~\eqref{Hauke} and perform first the one in $\omega$. Computing the inverse Fourier transform of $\tanh(\omega/2T)$ and using the fact that $\chi_{OO}(t,T)=2 \theta(t)\operatorname{Im} C_O(t,T)$ yields
\begin{equation}
F_Q(T)= -\frac{16k_B T}{\hbar}\int_{0}^\infty dt\, \operatorname{csch} \big(\pi k_B T\frac{t}{\hbar})\operatorname{Im} C_O.
\end{equation}
Therefore, using that $\int_{0}^{\infty} dx \, x\operatorname{csch} (\pi q x)=(8q^2)^{-1}$ $\forall q>0$, from our previous result \eqref{bound-ImCorr} we derive the following upper bound on the temperature-dependent quantum Fisher information
\begin{equation} \label{bound-Fish}
|F_Q(T)| \leq \frac{4}{k_BT} \braket{O\{ H-E_0,O\}},
\end{equation}
where $O$ is the operator that generates the transformation whose parameter is to be estimated, $H$ is the Hamiltonian of the system, and the expectation value is taken with respect to the corresponding thermal state at temperature $T$.

Let us finally note that, in this framework, the standard bounds \eqref{bound-susc1} and \eqref{bound-susc-Bog} given by the Heisenberg and Bogoliubov inequalities are divergent, due to the divergence of $ \operatorname{csch}(x)$ for $x\to 0$. Conversely, the time-linear dependence introduced by the QSL approach guarantees the convergence of the integral, yielding a finite upper bound on the quantum Fisher information. Making use of the celebrated Cramer-Rao bound $(\Delta \theta)^2 \geq (MF_Q)^{-1}$, Eq.~\eqref{bound-Fish} results in a lower bound on the variance of the parameter $\theta$ for $M$ independent measurements
\begin{equation}
(\Delta \theta)^2 \geq \frac{k_BT}{4} \frac{1}{\braket{O\{ H-E_0,O\}}},
\end{equation}
{which suggests that a better precision may be achieved at lower temperatures.}

\section{Discussion}
Conventional QSLs  identify the minimum time scale in which a process can unfold by exploiting the notion of quantum state distinguishability. 
Yet, many applications in theoretical and experimental physics are naturally formulated in terms of operator flows.  We have generalized the notion of QSL to this setting, providing bounds to the rate of unitary flows described by the conjugation of an observable by a one-parameter unitary.

Making use of Liouville space, we have derived analogs of the MT and ML QSLs, in which the minimum shift of the parameter required to distinguish the evolving operator from the initial one is lower bounded in terms of the mean and variance of the Liouvillian. These bounds generally exhibit a crossover, that we have characterized, and that is analogous to that observed in recent experiments for conventional QSLs.

We have also shown that QSLs for operator flows constrain the time dependence of autocorrelation functions and thus the dynamic susceptibilities introduced in linear response theory to describe transport coefficients. In the context of quantum parameter estimation, we have shown that QSLs for operator flows yield bounds on the quantum Fisher information that restricts the estimation error through the Cramer-Rao bound. This last application makes explicit the fact that the flow under consideration need not be on time, but can describe shifts of an arbitrary parameter through a continuous symmetry. The situation is thus analogous to the generalization of uncertainty relations for quantum states \cite{Braunstein96}.

Our results should find broad applications in nonequilibrium physics and, in particular, quantum technologies, including quantum metrology, quantum thermodynamics, and quantum computation. As we have demonstrated by several examples, our results are also of relevance in condensed matter physics to bound response functions and transport coefficients. We expect further applications of our results in other scenarios where operator flows naturally arise, such as the formulation of integrable systems in terms of Lax pairs and the Wegner renormalization group.

\section{Acknowledgements}
It is a pleasure to acknowledge discussions with Pablo Mart\'inez-Azcona, Norman Margolus,  Apollonas S. Matsoukas-Roubeas, Silvia Pappalardi, Giulio Cappelli and Jing Yang.

\appendix

\section{Deriving QSLs through a trigonometric approach}
\label{AppQSLTrig}

In this appendix, we show an alternative derivation that results in two weaker QSLs on the operator overlap, proportional to the tighter QSLs \eqref{QSL-ML} and \eqref{QSL-MT} derived in the main text. Making use of the trigonometric inequality
\begin{equation}
    \cos{x}\geq 1-\frac{2}{\pi}x-\frac{2}{\pi}\sin{x},
\end{equation}
valid for $x>0$, we obtain
\begin{equation*}
\begin{split}
    &\operatorname{Re} \braket{O_0|O_t}=\sum_{j>k}\cos(\frac{\Delta_{jk}t}{\hbar})\frac{(|O_{jk}|^2+|O_{kj}|^2)}{\|O\|^2}\geq \\
    &\sum_{j>k}\big(1-\frac{2}{\pi}\frac{\Delta_{jk}t}{\hbar}-\frac{2}{\pi}\sin{\frac{\Delta_{jk}t}{\hbar}}\big)\frac{(|O_{jk}|^2+|O_{kj}|^2)}{\|O\|^2}.
\end{split}
\end{equation*}
Therefore, being $\sin x\leq x$ for $x\geq0$, we find
\begin{equation}
    \operatorname{Re}\braket{O_0|O_t}\geq 1-\frac{4}{\pi}\braket{|\mathbb{L}|}t,
\end{equation}
that is,
\begin{equation} \label{QSL-ML_2}
    t \geq \frac{\pi}{4} \frac{1-\operatorname{Re}\braket{O_0|O_t}}{\braket{|\mathbb{L}|}}=\frac{\pi}{4} \frac{1-\cos{\mathcal{L}_t}}{\braket{|\mathbb{L}|}},
\end{equation}
which is proportional to the ML QSL \eqref{QSL-ML} derived in the main text through a constant smaller than one. Next, let us make use of the trigonometric inequality
\begin{equation}
    \cos{x}\geq 1-\frac{4}{\pi^2}x\sin{x}-\frac{2}{\pi^2}x^2,
\end{equation}
which holds again $\forall x$. Combining it with the bound \eqref{O'upperbound2} on $\operatorname{Re}\braket{O_0|\dot{O}_t}$, we obtain
\begin{equation*}    
\begin{split}           
    &\operatorname{Re}\braket{O_0|O_t}\geq\sum_{j,k}
    \frac{|O_{jk}|^2}{\|O\|^2}[1-\frac{4}{\pi^2}\frac{\Delta_{jk}t}{\hbar}\sin{\frac{\Delta_{jk}t}{\hbar}}-\\
    &\frac{2}{\pi^2}(\frac{\Delta_{jk}t}{\hbar})^2]=1+\frac{4}{\pi^2}\operatorname{Re} \braket{O_0|\dot{O}_t}t-\frac{2}{\pi^2}(\Delta\mathbb{L})^2t^2\\
    &\geq 1-\frac{6}{\pi^2}(\Delta\mathbb{L})^2t^2,
\end{split}
\end{equation*}
that is,
\begin{equation} \label{QSL-MT_2}
    t \geq \frac{\pi}{\sqrt{6}} \frac{\sqrt{1-\operatorname{Re}\braket{O_0|O_t}}}{\Delta\mathbb{L}} =\frac{\pi}{\sqrt{6}} \frac{\sqrt{1- \cos{\mathcal{L}_t}}} {\Delta\mathbb{L}}.
\end{equation}
As for the case of ML, this bound is proportional to our MT QSL \eqref{QSL-MT} for operators through a constant smaller than one, thus yielding a weaker result.

\section{QSLs in the Schr{\"o}dinger picture}\label{AppQSLSchro}

In this appendix, we reformulate our results on operator flows in the context of the standard time-evolution of quantum states and compare them with the well-known MT \cite{Mandelstam45} and ML \cite{Margolus98} bounds. In this way, we provide a further justification for the bounds \eqref{QSL-MT} and \eqref{QSL-ML} to be regarded as generalizations of the MT and ML quantum speed limits to operator flows, respectively. To this end, let us choose the initial operator $O_0$ to be the projector onto the pure state $\ket{\psi_0}$, $O_0=\ket{\psi_0}\bra{\psi_0}$, and let the expansion of $\ket{\psi_0}$ in the energy eigenbasis be
\begin{equation} \label{state-dec}
    \ket{\psi_0}=\sum_j c_j\ket{j},
\end{equation}
where $H\ket{j}=E_j\ket{j}$ and $\sum_j|c_j|^2=1$. Then, the vectorization of $O_0$ is $\ket{O_0}=\sum_{jk}c^*_jc_k\ket{j}\ket{k}$, with $\|O_0\|=1$. Since $\ket{\psi_t}=U_t\ket{\psi_0}$, where $U_t=e^{-iHt/\hbar}$, in order to reproduce the forward time-evolution of the quantum state we need to evolve the operator backward in time:
\begin{equation}
    O_{-t}=U^\dagger_{-t}O_0U_{-t}=U_t\ket{\psi_0}\bra{\psi_0}U^\dagger_t=\ket{\psi_t}\bra{\psi_t}.
\end{equation}
The operator overlap then reduces to the square modulus of the state overlap, that is, to the Uhlmann fidelity
\begin{equation}
    \braket{O_0|O_{-t}}=\Tr{\left(O_0^\dagger O_{-t}\right)}=|\braket{\psi_0|\psi_t}|^2.
\end{equation}
Moreover, being the Bures angle between states defined as
\begin{equation}
    \ell_t=\arccos|\braket{\psi_0|\psi_t}|,
\end{equation}
we have that $1-\braket{O_0|O_{-t}}=\sin^2{\ell_t}$. We observe that the operator angle $\mathcal{L}_t$ defined in Eq.~\eqref{defLt} does not reduce to the standard Bures angle between states, being $\mathcal{L}_{-t}=\arccos{|\braket{\psi_0|\psi_t}|^2}$ when $O_0=\ket{\psi_0}\bra{\psi_0}$.

The typical maximal speeds of the flow, given by the ML \eqref{QSL-ML} and MT \eqref{QSL-MT} QSLs respectively, can be written as
\begin{equation}
\begin{split}
    \braket{|\mathbb{L}|}&=\frac{1}{\hbar}\sum_{jk}|\Delta_{jk}||c_j|^2|c_k|^2,\\ (\Delta\mathbb{L})^2&=\frac{1}{\hbar^2}\sum_{jk}\Delta^2_{jk}|c_j|^2|c_k|^2.
\end{split}
\end{equation}
From the second equation, being also
\begin{equation}
    (\Delta H)^2=\sum_jE_j^2|c_j|^2 -\sum_{jk}E_jE_k|c_j|^2|c_k|^2,
\end{equation}
we derive a proportionality relation between the Hamiltonian and Liouvillian variances (over $\ket{\psi_0}$ and $\ket{O_0}$, respectively):
\begin{equation}
    (\Delta\mathbb{L})^2=2\frac{(\Delta H)^2}{\hbar^2}.
\end{equation}
Therefore, the MT bound \eqref{QSL-MT} derived in the main text for operators reduces to the following QSL for states, given in terms of the energy variance:
\begin{equation}
    t\geq \frac{ \hbar}{\Delta H} \sin{\ell_t}.
\end{equation}
In the case of orthogonal evolution $\braket{\psi_0|\psi_\tau}=0$, we have $\sin{\ell_\tau}=1$ and therefore we obtain a bound proportional to (though weaker than) the standard Mandelstam-Tamm QSL \cite{Mandelstam45},
\begin{equation}
    \tau \geq \tau_{MT}=\frac{\pi\hbar}{2\Delta H}>\frac{ \hbar}{\Delta H},
\end{equation}
thus justifying the interpretation of Eq.~\eqref{QSL-MT} as an MT-type of QSL for operators.

On the other hand, the mean energy can be rewritten as
\begin{equation}
    \braket{H}=\sum_jE_j|c_j|^2=\frac{1}{2}\sum_{jk}(E_j+E_k)|c_j|^2|c_k|^2,
\end{equation}
which, by using that $|\Delta_{jk}|=|E_j-E_k|\leq E_j+ E_k-2E_0$, implies
\begin{equation}\label{C1mean-ene}
    \braket{|\mathbb{L}|}\leq 2\frac{\braket{H}-E_0}{\hbar}.
\end{equation}
Therefore, we can recast the bound \eqref{QSL-ML} as
\begin{equation} \label{QSL-ML-states}
    t\geq \frac{ \hbar}{2\alpha} \frac{\sin^2{\ell_t}}{\braket{H}-E_0},
\end{equation}
which represents a generalization of the ML QSL \cite{Margolus98} to the case of an arbitrary angle between the initial and final states. A similar generalization was also claimed in reference \cite{Deffner2013JPA}, though their bound is not given in terms of $\braket{H}-E_0$, but rather in terms of the quantity $|\braket{H}|$.  To our knowledge, the generalization of the ML bound to arbitrary angles between initial and final states was only proven numerically \cite{Giovannetti2003,Giovannetti2004,Deffner2017}. Let us note that, by considering again the orthogonalization time $t=\tau$, our QSL becomes
\begin{equation}
    \tau \geq \tau_{ML}>\frac{ \hbar}{2\alpha(\braket{H}-E_0)}\approx 0.44\,\tau_{ML},
\end{equation}
where $\tau_{ML}={\pi\hbar}/{2( \braket{H}-E_0)}$ is the well known ML QSL \cite{Margolus98} for orthogonal evolution.

In conclusion, our bounds \eqref{QSL-ML} and \eqref{QSL-MT} on operator flows turn out to be proportional to the standard MT and ML QSLs for state evolution, justifying the given interpretation. The constant of proportionality is smaller than one, meaning that our bounds are not violated, though they are not tight for quantum state evolution. 

\begin{figure}[t!]
	\centering
	\includegraphics[width=0.45\textwidth]{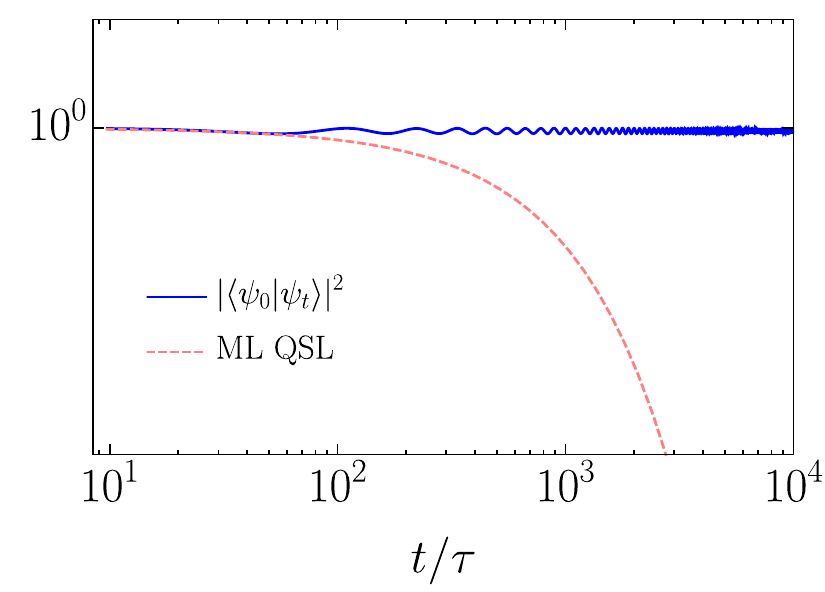}
	\caption{The coherent Gibbs state overlap undergoes an initial decay which is qualitatively captured by the ML QSL for states \eqref{QSL-ML-overlap}. The parameters chosen for the simulation are $\sigma=1$, $d=50$, and inverse temperature $\beta=10$.}
	\label{fig-RMT}
\end{figure}
Finally, let us perform a numerical test of our ML QSL for states \eqref{QSL-ML-states}, which can be recast as a lower bound on the state overlap
\begin{equation} \label{QSL-ML-overlap}
|\braket{\psi_0|\psi_t}|^2 \geq 1- 2\alpha(\braket{H}-E_0)t/\hbar.
\end{equation}
In Fig.~\ref{fig-RMT}, we test the bound \eqref{QSL-ML-overlap} for a single realization of a random matrix Hamiltonian $H$ of dimension $d=50$, generated from the Gaussian Orthogonal Ensemble (GOE) with variance $\sigma=1$. The initial state is chosen to be the coherent Gibbs state at inverse temperature $\beta$
\begin{equation}
\ket{\psi_0}=\frac{1}{\sqrt{Z}}\sum_{n=1}^{d} e^{-\beta\frac{E_n}{2}} \ket{n} ,
\end{equation}
where $Z=\sum_ne^{-\beta E_n}$ is the partition function. In this setting, the quantum state overlap can be conveniently rewritten in terms of the analytically-continued partition function as follows \cite{Delcampo17,Xu21}
\begin{equation}
\braket{\psi_0|\psi_t}= \frac{Z(\beta+it)}{Z(\beta)}.
\end{equation}
After the initial decay shown in Fig.~\ref{fig-RMT}, the fidelity undergoes an oscillatory behavior that is no longer captured by the QSL \eqref{QSL-ML-overlap}. Deviations from the bound appear far after the characteristic timescale $\tau=(\sigma\sqrt{8d})^{-1}$, where $\sigma\sqrt{8d}$ correspond to the width of the eigenvalue distribution (i.e.,~to the diameter of the Wigner semicircle law), that is to the largest frequency involved in the evolution. 

\section{Margolous-Levitin QSL on operator flows with driven Hamiltonians}\label{MLOpFlowDriven}

Finally, we partially extend our results to the case in which the Hamiltonian generating the unitary flow $O_t=U^\dagger_tO_0U_t$ is time-dependent, that is
\begin{equation}
    U_t=\mathcal{T}\exp{-\frac{i}{\hbar}\int_{0}^{t}H(s) \,ds}.
\end{equation}
We focus on the case in which $O_0$ is an observable, i.e., a Hermitian operator. Moreover, let us assume that the Hamiltonians at different times commute so that we can drop the time-ordering operator $\mathcal{T}$ in front of the exponential. In addition, if we make the strongest assumption that $H_t$ remains diagonal in the initial eigenbasis,
\begin{equation}
    H_t=\sum_j E_j(t) \ket{j}\bra{j},
\end{equation}
where only the eigenvalues $E_j(t)$ depend on time, then we can write the action of $U_t$ over the energy eigenvector $\ket{j}$ as
\begin{equation} 
    U_t \ket{j}=e^{-\frac{i}{\hbar}\int_{0}^{t}E_j(s)ds}\ket{j}.
\end{equation}
Under this assumption, the operator overlap can be expanded as
\begin{equation}
    \braket{O_0|O_t}=\frac{1}{\|O\|^2}\sum_{j,k}e^{i\frac{\overline{\Delta_{jk}}(t)}{\hbar}t}|O_{jk}|^2,
\end{equation}
so that, making use of the inequality $\cos x \geq 1-\alpha |x|$ \cite{Andersson2019} as in the main text, we derive an ML bound on operator flows in the case of driven dynamics,
\begin{equation} \label{QSL-ML-t}
    t \geq\frac{1-\operatorname{Re}\braket{O_0|O_t}}{\alpha\overline{\braket{|\mathbb{L}|}}}=\frac{1-\cos{\mathcal{L}_t}}{\alpha\overline{\braket{|\mathbb{L}|}}},
\end{equation}
where $\overline{f}(t)=\frac{1}{t}\int_{0}^{t} f(u) du$ is the time average at time $t$. As a consequence, by choosing $O_0=\ket{\psi_0}\bra{\psi_0}$, we obtain a generalization of the ML QSL for state evolution under driven Hamiltonians and arbitrary angles:
\begin{equation} 
    t\geq \frac{ \hbar}{2\alpha}\frac{\sin^2{\ell_t}}{\overline{\braket{H}-E_0}},
\end{equation}
valid when the energy eigenvectors are stationary. A related result was also claimed in reference \cite{Deffner2013PRL}, but their derivation was later shown to generalize the MT QSL rather than the ML one, as pointed out in \cite{Okuyama2018Comment}.

\medskip

\bibliography{QSL_biblio}

\begin{thebibliography}{81}
\providecommand{\natexlab}[1]{#1}
\providecommand{\url}[1]{\texttt{#1}}
\expandafter\ifx\csname urlstyle\endcsname\relax
  \providecommand{\doi}[1]{doi: #1}\else
  \providecommand{\doi}{doi: \begingroup \urlstyle{rm}\Url}\fi

\bibitem[Mandelstam and Tamm(1945)]{Mandelstam45}
L.~Mandelstam and I.~Tamm.
\newblock The uncertainty relation between energy and time in non-relativistic
  quantum mechanics.
\newblock \emph{J. Phys. USSR}, 9:\penalty0 249, 1945.
\newblock \doi{https://doi.org/10.1007/978-3-642-74626-0_8}.

\bibitem[Margolus and Levitin(1998)]{Margolus98}
Norman Margolus and Lev~B. Levitin.
\newblock The maximum speed of dynamical evolution.
\newblock \emph{Physica D: Nonlinear Phenomena}, 120\penalty0 (1):\penalty0
  188--195, 1998.
\newblock ISSN 0167-2789.
\newblock \doi{https://doi.org/10.1016/S0167-2789(98)00054-2}.
\newblock URL
  \url{https://www.sciencedirect.com/science/article/pii/S0167278998000542}.
\newblock Proceedings of the Fourth Workshop on Physics and Consumption.

\bibitem[Uhlmann(1992)]{Uhlmann1992}
Armin Uhlmann.
\newblock An energy dispersion estimate.
\newblock \emph{Physics Letters A}, 161\penalty0 (4):\penalty0 329 -- 331,
  1992.
\newblock ISSN 0375-9601.
\newblock \doi{https://doi.org/10.1016/0375-9601(92)90555-Z}.
\newblock URL
  \url{http://www.sciencedirect.com/science/article/pii/037596019290555Z}.

\bibitem[Campaioli et~al.(2018)Campaioli, Pollock, Binder, and
  Modi]{Campaioli2018}
Francesco Campaioli, Felix~A. Pollock, Felix~C. Binder, and Kavan Modi.
\newblock Tightening quantum speed limits for almost all states.
\newblock \emph{Phys. Rev. Lett.}, 120:\penalty0 060409, Feb 2018.
\newblock \doi{10.1103/PhysRevLett.120.060409}.
\newblock URL \url{https://link.aps.org/doi/10.1103/PhysRevLett.120.060409}.

\bibitem[Anandan and Aharonov(1990)]{Anandan1990}
J.~Anandan and Y.~Aharonov.
\newblock Geometry of quantum evolution.
\newblock \emph{Phys. Rev. Lett.}, 65:\penalty0 1697--1700, Oct 1990.
\newblock \doi{10.1103/PhysRevLett.65.1697}.
\newblock URL \url{https://link.aps.org/doi/10.1103/PhysRevLett.65.1697}.

\bibitem[Deffner and Lutz(2013{\natexlab{a}})]{Deffner2013JPA}
Sebastian Deffner and Eric Lutz.
\newblock Energy{\textendash}time uncertainty relation for driven quantum
  systems.
\newblock \emph{Journal of Physics A: Mathematical and Theoretical},
  46\penalty0 (33):\penalty0 335302, jul 2013{\natexlab{a}}.
\newblock \doi{10.1088/1751-8113/46/33/335302}.
\newblock URL \url{https://doi.org/10.1088/1751-8113/46/33/335302}.

\bibitem[Okuyama and Ohzeki(2018{\natexlab{a}})]{Okuyama18}
Manaka Okuyama and Masayuki Ohzeki.
\newblock Comment on `energy-time uncertainty relation for driven quantum
  systems'.
\newblock \emph{Journal of Physics A: Mathematical and Theoretical},
  51\penalty0 (31):\penalty0 318001, jun 2018{\natexlab{a}}.
\newblock \doi{10.1088/1751-8121/aacb90}.
\newblock URL \url{https://doi.org/10.1088/1751-8121/aacb90}.

\bibitem[Taddei et~al.(2013)Taddei, Escher, Davidovich, and
  de~Matos~Filho]{Taddei2013}
M.~M. Taddei, B.~M. Escher, L.~Davidovich, and R.~L. de~Matos~Filho.
\newblock Quantum speed limit for physical processes.
\newblock \emph{Phys. Rev. Lett.}, 110:\penalty0 050402, Jan 2013.
\newblock \doi{10.1103/PhysRevLett.110.050402}.
\newblock URL \url{https://link.aps.org/doi/10.1103/PhysRevLett.110.050402}.

\bibitem[del Campo et~al.(2013)del Campo, Egusquiza, Plenio, and
  Huelga]{Delcampo2013}
A.~del Campo, I.~L. Egusquiza, M.~B. Plenio, and S.~F. Huelga.
\newblock Quantum speed limits in open system dynamics.
\newblock \emph{Phys. Rev. Lett.}, 110:\penalty0 050403, Jan 2013.
\newblock \doi{10.1103/PhysRevLett.110.050403}.
\newblock URL \url{https://link.aps.org/doi/10.1103/PhysRevLett.110.050403}.

\bibitem[Deffner and Lutz(2013{\natexlab{b}})]{Deffner2013PRL}
Sebastian Deffner and Eric Lutz.
\newblock Quantum speed limit for non-markovian dynamics.
\newblock \emph{Phys. Rev. Lett.}, 111:\penalty0 010402, Jul
  2013{\natexlab{b}}.
\newblock \doi{10.1103/PhysRevLett.111.010402}.
\newblock URL \url{https://link.aps.org/doi/10.1103/PhysRevLett.111.010402}.

\bibitem[Campaioli et~al.(2019)Campaioli, Pollock, and Modi]{Campaioli2019}
Francesco Campaioli, Felix~A. Pollock, and Kavan Modi.
\newblock Tight, robust, and feasible quantum speed limits for open dynamics.
\newblock \emph{{Quantum}}, 3:\penalty0 168, August 2019.
\newblock ISSN 2521-327X.
\newblock \doi{10.22331/q-2019-08-05-168}.
\newblock URL \url{https://doi.org/10.22331/q-2019-08-05-168}.

\bibitem[Garc{\'{\i}}a-Pintos and del Campo(2019)]{GarciaPintos19}
Luis~Pedro Garc{\'{\i}}a-Pintos and Adolfo del Campo.
\newblock Quantum speed limits under continuous quantum measurements.
\newblock \emph{New Journal of Physics}, 21\penalty0 (3):\penalty0 033012, mar
  2019.
\newblock \doi{10.1088/1367-2630/ab099e}.
\newblock URL \url{https://doi.org/10.1088/1367-2630/ab099e}.

\bibitem[Shanahan et~al.(2018)Shanahan, Chenu, Margolus, and del
  Campo]{Shanahan2018}
B.~Shanahan, A.~Chenu, N.~Margolus, and A.~del Campo.
\newblock Quantum speed limits across the quantum-to-classical transition.
\newblock \emph{Phys. Rev. Lett.}, 120:\penalty0 070401, Feb 2018.
\newblock \doi{10.1103/PhysRevLett.120.070401}.
\newblock URL \url{https://link.aps.org/doi/10.1103/PhysRevLett.120.070401}.

\bibitem[Okuyama and Ohzeki(2018{\natexlab{b}})]{Okuyama2018PRL}
Manaka Okuyama and Masayuki Ohzeki.
\newblock Quantum speed limit is not quantum.
\newblock \emph{Phys. Rev. Lett.}, 120:\penalty0 070402, Feb
  2018{\natexlab{b}}.
\newblock \doi{10.1103/PhysRevLett.120.070402}.
\newblock URL \url{https://link.aps.org/doi/10.1103/PhysRevLett.120.070402}.

\bibitem[Shiraishi et~al.(2018)Shiraishi, Funo, and Saito]{Shiraishi18}
Naoto Shiraishi, Ken Funo, and Keiji Saito.
\newblock Speed limit for classical stochastic processes.
\newblock \emph{Phys. Rev. Lett.}, 121:\penalty0 070601, Aug 2018.
\newblock \doi{10.1103/PhysRevLett.121.070601}.
\newblock URL \url{https://link.aps.org/doi/10.1103/PhysRevLett.121.070601}.

\bibitem[Deffner and Campbell(2017)]{Deffner2017}
Sebastian Deffner and Steve Campbell.
\newblock Quantum speed limits: from heisenberg's uncertainty principle to
  optimal quantum control.
\newblock \emph{Journal of Physics A: Mathematical and Theoretical},
  50\penalty0 (45):\penalty0 453001, oct 2017.
\newblock \doi{10.1088/1751-8121/aa86c6}.
\newblock URL \url{https://doi.org/10.1088/1751-8121/aa86c6}.

\bibitem[Lloyd(2000)]{Lloyd2000}
S.~Lloyd.
\newblock Ultimate physical limits to computation.
\newblock \emph{Nature}, 406\penalty0 (6799):\penalty0 1047--1054, 2000.
\newblock \doi{https://doi.org/10.1038/35023282}.

\bibitem[Lloyd(2002)]{Lloyd2002}
Seth Lloyd.
\newblock Computational capacity of the universe.
\newblock \emph{Phys. Rev. Lett.}, 88:\penalty0 237901, May 2002.
\newblock \doi{10.1103/PhysRevLett.88.237901}.
\newblock URL \url{https://link.aps.org/doi/10.1103/PhysRevLett.88.237901}.

\bibitem[Giovannetti et~al.(2011)Giovannetti, Lloyd, and
  Maccone]{Giovannetti2011}
Vittorio Giovannetti, Seth Lloyd, and Lorenzo Maccone.
\newblock Advances in quantum metrology.
\newblock \emph{Nature Photonics}, 5\penalty0 (4):\penalty0 222--229, 2011.
\newblock ISSN 1749-4893.
\newblock \doi{10.1038/nphoton.2011.35}.
\newblock URL \url{https://doi.org/10.1038/nphoton.2011.35}.

\bibitem[Beau and del Campo(2017)]{Beau17}
M.~Beau and A.~del Campo.
\newblock Nonlinear quantum metrology of many-body open systems.
\newblock \emph{Phys. Rev. Lett.}, 119:\penalty0 010403, Jul 2017.
\newblock \doi{10.1103/PhysRevLett.119.010403}.
\newblock URL \url{https://link.aps.org/doi/10.1103/PhysRevLett.119.010403}.

\bibitem[Caneva et~al.(2009)Caneva, Murphy, Calarco, Fazio, Montangero,
  Giovannetti, and Santoro]{Caneva2009}
T.~Caneva, M.~Murphy, T.~Calarco, R.~Fazio, S.~Montangero, V.~Giovannetti, and
  G.~E. Santoro.
\newblock Optimal control at the quantum speed limit.
\newblock \emph{Phys. Rev. Lett.}, 103:\penalty0 240501, Dec 2009.
\newblock \doi{10.1103/PhysRevLett.103.240501}.
\newblock URL \url{https://link.aps.org/doi/10.1103/PhysRevLett.103.240501}.

\bibitem[Hegerfeldt(2013)]{Hegerfeldt13}
Gerhard~C. Hegerfeldt.
\newblock Driving at the quantum speed limit: Optimal control of a two-level
  system.
\newblock \emph{Phys. Rev. Lett.}, 111:\penalty0 260501, Dec 2013.
\newblock \doi{10.1103/PhysRevLett.111.260501}.
\newblock URL \url{https://link.aps.org/doi/10.1103/PhysRevLett.111.260501}.

\bibitem[Funo et~al.(2017)Funo, Zhang, Chatou, Kim, Ueda, and del
  Campo]{Funo17}
Ken Funo, Jing-Ning Zhang, Cyril Chatou, Kihwan Kim, Masahito Ueda, and Adolfo
  del Campo.
\newblock Universal work fluctuations during shortcuts to adiabaticity by
  counterdiabatic driving.
\newblock \emph{Phys. Rev. Lett.}, 118:\penalty0 100602, Mar 2017.
\newblock \doi{10.1103/PhysRevLett.118.100602}.
\newblock URL \url{https://link.aps.org/doi/10.1103/PhysRevLett.118.100602}.

\bibitem[Campbell and Deffner(2017)]{Campbell2017}
Steve Campbell and Sebastian Deffner.
\newblock Trade-off between speed and cost in shortcuts to adiabaticity.
\newblock \emph{Phys. Rev. Lett.}, 118:\penalty0 100601, Mar 2017.
\newblock \doi{10.1103/PhysRevLett.118.100601}.
\newblock URL \url{https://link.aps.org/doi/10.1103/PhysRevLett.118.100601}.

\bibitem[Alipour et~al.(2020)Alipour, Chenu, Rezakhani, and del
  Campo]{Alipour20}
Sahar Alipour, Aurelia Chenu, Ali~T. Rezakhani, and Adolfo del Campo.
\newblock Shortcuts to {A}diabaticity in {D}riven {O}pen {Q}uantum {S}ystems:
  {B}alanced {G}ain and {L}oss and {N}on-{M}arkovian {E}volution.
\newblock \emph{{Quantum}}, 4:\penalty0 336, September 2020.
\newblock ISSN 2521-327X.
\newblock \doi{10.22331/q-2020-09-28-336}.
\newblock URL \url{https://doi.org/10.22331/q-2020-09-28-336}.

\bibitem[Funo et~al.(2021)Funo, Lambert, and Nori]{Funo21}
Ken Funo, Neill Lambert, and Franco Nori.
\newblock General bound on the performance of counter-diabatic driving acting
  on dissipative spin systems.
\newblock \emph{Phys. Rev. Lett.}, 127:\penalty0 150401, Oct 2021.
\newblock \doi{10.1103/PhysRevLett.127.150401}.
\newblock URL \url{https://link.aps.org/doi/10.1103/PhysRevLett.127.150401}.

\bibitem[Bukov et~al.(2019)Bukov, Sels, and Polkovnikov]{Bukov2019}
Marin Bukov, Dries Sels, and Anatoli Polkovnikov.
\newblock Geometric speed limit of accessible many-body state preparation.
\newblock \emph{Phys. Rev. X}, 9:\penalty0 011034, Feb 2019.
\newblock \doi{10.1103/PhysRevX.9.011034}.
\newblock URL \url{https://link.aps.org/doi/10.1103/PhysRevX.9.011034}.

\bibitem[Suzuki and Takahashi(2020)]{Suzuki20}
Keisuke Suzuki and Kazutaka Takahashi.
\newblock Performance evaluation of adiabatic quantum computation via quantum
  speed limits and possible applications to many-body systems.
\newblock \emph{Phys. Rev. Research}, 2:\penalty0 032016, Jul 2020.
\newblock \doi{10.1103/PhysRevResearch.2.032016}.
\newblock URL \url{https://link.aps.org/doi/10.1103/PhysRevResearch.2.032016}.

\bibitem[del Campo(2021)]{Delcampo21}
Adolfo del Campo.
\newblock Probing quantum speed limits with ultracold gases.
\newblock \emph{Phys. Rev. Lett.}, 126:\penalty0 180603, May 2021.
\newblock \doi{10.1103/PhysRevLett.126.180603}.
\newblock URL \url{https://link.aps.org/doi/10.1103/PhysRevLett.126.180603}.

\bibitem[Hamazaki(2022)]{Hamazaki2021}
Ryusuke Hamazaki.
\newblock Speed limits for macroscopic transitions.
\newblock \emph{PRX Quantum}, 3:\penalty0 020319, Apr 2022.
\newblock \doi{10.1103/PRXQuantum.3.020319}.
\newblock URL \url{https://link.aps.org/doi/10.1103/PRXQuantum.3.020319}.

\bibitem[Gong and Hamazaki(2022)]{Hamazaki2022}
Zongping Gong and Ryusuke Hamazaki.
\newblock Bounds in nonequilibrium quantum dynamics.
\newblock \emph{International Journal of Modern Physics B}, 36\penalty0
  (31):\penalty0 2230007, 2022.
\newblock \doi{10.1142/S0217979222300079}.
\newblock URL \url{https://doi.org/10.1142/S0217979222300079}.

\bibitem[Jing et~al.(2016)Jing, Wu, and del Campo]{Jing16}
Jun Jing, Lian-Ao Wu, and Adolfo del Campo.
\newblock Fundamental speed limits to the generation of quantumness.
\newblock \emph{Scientific Reports}, 6\penalty0 (1):\penalty0 38149, Nov 2016.
\newblock ISSN 2045-2322.
\newblock \doi{10.1038/srep38149}.
\newblock URL \url{https://doi.org/10.1038/srep38149}.

\bibitem[Marvian et~al.(2016)Marvian, Spekkens, and Zanardi]{Marvian16}
Iman Marvian, Robert~W. Spekkens, and Paolo Zanardi.
\newblock Quantum speed limits, coherence, and asymmetry.
\newblock \emph{Phys. Rev. A}, 93:\penalty0 052331, May 2016.
\newblock \doi{10.1103/PhysRevA.93.052331}.
\newblock URL \url{https://link.aps.org/doi/10.1103/PhysRevA.93.052331}.

\bibitem[Mohan et~al.(2022)Mohan, Das, and Pati]{Mohan22}
Brij Mohan, Siddhartha Das, and Arun~Kumar Pati.
\newblock Quantum speed limits for information and coherence.
\newblock \emph{New Journal of Physics}, 24\penalty0 (6):\penalty0 065003, jun
  2022.
\newblock \doi{10.1088/1367-2630/ac753c}.
\newblock URL \url{https://doi.org/10.1088/1367-2630/ac753c}.

\bibitem[Campaioli et~al.(2022)Campaioli, shui Yu, Pollock, and
  Modi]{Campaioli22}
Francesco Campaioli, Chang shui Yu, Felix~A Pollock, and Kavan Modi.
\newblock Resource speed limits: maximal rate of resource variation.
\newblock \emph{New Journal of Physics}, 24\penalty0 (6):\penalty0 065001, jun
  2022.
\newblock \doi{10.1088/1367-2630/ac7346}.
\newblock URL \url{https://doi.org/10.1088/1367-2630/ac7346}.

\bibitem[Gingrich et~al.(2016)Gingrich, Horowitz, Perunov, and
  England]{Gingrich2016}
Todd~R. Gingrich, Jordan~M. Horowitz, Nikolay Perunov, and Jeremy~L. England.
\newblock Dissipation bounds all steady-state current fluctuations.
\newblock \emph{Phys. Rev. Lett.}, 116:\penalty0 120601, Mar 2016.
\newblock \doi{10.1103/PhysRevLett.116.120601}.
\newblock URL \url{https://link.aps.org/doi/10.1103/PhysRevLett.116.120601}.

\bibitem[Hasegawa(2021)]{Hasegawa2021}
Yoshihiko Hasegawa.
\newblock Thermodynamic uncertainty relation for general open quantum systems.
\newblock \emph{Phys. Rev. Lett.}, 126:\penalty0 010602, Jan 2021.
\newblock \doi{10.1103/PhysRevLett.126.010602}.
\newblock URL \url{https://link.aps.org/doi/10.1103/PhysRevLett.126.010602}.

\bibitem[Nicholson et~al.(2020)Nicholson, Garc{\'\i}a-Pintos, del Campo, and
  Green]{Nicholson2020}
Schuyler~B. Nicholson, Luis~Pedro Garc{\'\i}a-Pintos, Adolfo del Campo, and
  Jason~R. Green.
\newblock Time--information uncertainty relations in thermodynamics.
\newblock \emph{Nature Physics}, 16\penalty0 (12):\penalty0 1211--1215, Dec
  2020.
\newblock ISSN 1745-2481.
\newblock \doi{10.1038/s41567-020-0981-y}.
\newblock URL \url{https://doi.org/10.1038/s41567-020-0981-y}.

\bibitem[Vo et~al.(2020)Vo, Van~Vu, and Hasegawa]{Vo20}
Van~Tuan Vo, Tan Van~Vu, and Yoshihiko Hasegawa.
\newblock Unified approach to classical speed limit and thermodynamic
  uncertainty relation.
\newblock \emph{Phys. Rev. E}, 102:\penalty0 062132, Dec 2020.
\newblock \doi{10.1103/PhysRevE.102.062132}.
\newblock URL \url{https://link.aps.org/doi/10.1103/PhysRevE.102.062132}.

\bibitem[Garc\'{\i}a-Pintos et~al.(2022)Garc\'{\i}a-Pintos, Nicholson, Green,
  del Campo, and Gorshkov]{Garcia2022}
Luis~Pedro Garc\'{\i}a-Pintos, Schuyler~B. Nicholson, Jason~R. Green, Adolfo
  del Campo, and Alexey~V. Gorshkov.
\newblock Unifying quantum and classical speed limits on observables.
\newblock \emph{Phys. Rev. X}, 12:\penalty0 011038, Feb 2022.
\newblock \doi{10.1103/PhysRevX.12.011038}.
\newblock URL \url{https://link.aps.org/doi/10.1103/PhysRevX.12.011038}.

\bibitem[Mohan and Pati(2022)]{Mohan21}
Brij Mohan and Arun~Kumar Pati.
\newblock Quantum speed limits for observables.
\newblock \emph{Phys. Rev. A}, 106:\penalty0 042436, Oct 2022.
\newblock \doi{10.1103/PhysRevA.106.042436}.
\newblock URL \url{https://link.aps.org/doi/10.1103/PhysRevA.106.042436}.

\bibitem[Perelomov(1990)]{Perelomov1989integrable}
A.M. Perelomov.
\newblock \emph{Integrable Systems of Classical Mechanics and Lie Algebras
  Volume I}.
\newblock Birkhäuser Basel, 1990.
\newblock \doi{https://doi.org/10.1007/978-3-0348-9257-5}.

\bibitem[Wegner(2001)]{Wegner2001}
Franz~J. Wegner.
\newblock Flow equations for hamiltonians.
\newblock \emph{Physics Reports}, 348\penalty0 (1):\penalty0 77--89, 2001.
\newblock ISSN 0370-1573.
\newblock \doi{https://doi.org/10.1016/S0370-1573(00)00136-8}.
\newblock URL
  \url{https://www.sciencedirect.com/science/article/pii/S0370157300001368}.

\bibitem[Poggi(2019)]{Poggi2019}
Pablo~M. Poggi.
\newblock Geometric quantum speed limits and short-time accessibility to
  unitary operations.
\newblock \emph{Phys. Rev. A}, 99:\penalty0 042116, Apr 2019.
\newblock \doi{10.1103/PhysRevA.99.042116}.
\newblock URL \url{https://link.aps.org/doi/10.1103/PhysRevA.99.042116}.

\bibitem[Uzdin(2013)]{Uzdin2013}
Raam Uzdin.
\newblock Resources needed for non-unitary quantum operations.
\newblock \emph{Journal of Physics A: Mathematical and Theoretical},
  46\penalty0 (14):\penalty0 145302, mar 2013.
\newblock \doi{10.1088/1751-8113/46/14/145302}.
\newblock URL \url{https://doi.org/10.1088%2F1751-8113%2F46%2F14%2F145302}.

\bibitem[Uzdin and Kosloff(2016)]{Uzdin2016}
Raam Uzdin and Ronnie Kosloff.
\newblock Speed limits in liouville space for open quantum systems.
\newblock \emph{{EPL} (Europhysics Letters)}, 115\penalty0 (4):\penalty0 40003,
  aug 2016.
\newblock \doi{10.1209/0295-5075/115/40003}.
\newblock URL \url{https://doi.org/10.1209/0295-5075/115/40003}.

\bibitem[von Keyserlingk et~al.(2018)von Keyserlingk, Rakovszky, Pollmann, and
  Sondhi]{Keyserlingk2018}
C.~W. von Keyserlingk, Tibor Rakovszky, Frank Pollmann, and S.~L. Sondhi.
\newblock Operator hydrodynamics, otocs, and entanglement growth in systems
  without conservation laws.
\newblock \emph{Phys. Rev. X}, 8:\penalty0 021013, Apr 2018.
\newblock \doi{10.1103/PhysRevX.8.021013}.
\newblock URL \url{https://link.aps.org/doi/10.1103/PhysRevX.8.021013}.

\bibitem[Khemani et~al.(2018)Khemani, Vishwanath, and Huse]{Khemani2018}
Vedika Khemani, Ashvin Vishwanath, and David~A. Huse.
\newblock Operator spreading and the emergence of dissipative hydrodynamics
  under unitary evolution with conservation laws.
\newblock \emph{Phys. Rev. X}, 8:\penalty0 031057, Sep 2018.
\newblock \doi{10.1103/PhysRevX.8.031057}.
\newblock URL \url{https://link.aps.org/doi/10.1103/PhysRevX.8.031057}.

\bibitem[Nahum et~al.(2018)Nahum, Vijay, and Haah]{Nahum2018}
Adam Nahum, Sagar Vijay, and Jeongwan Haah.
\newblock Operator spreading in random unitary circuits.
\newblock \emph{Phys. Rev. X}, 8:\penalty0 021014, Apr 2018.
\newblock \doi{10.1103/PhysRevX.8.021014}.
\newblock URL \url{https://link.aps.org/doi/10.1103/PhysRevX.8.021014}.

\bibitem[Gopalakrishnan et~al.(2018)Gopalakrishnan, Huse, Khemani, and
  Vasseur]{Gopalakrishnan2018}
Sarang Gopalakrishnan, David~A. Huse, Vedika Khemani, and Romain Vasseur.
\newblock Hydrodynamics of operator spreading and quasiparticle diffusion in
  interacting integrable systems.
\newblock \emph{Phys. Rev. B}, 98:\penalty0 220303, Dec 2018.
\newblock \doi{10.1103/PhysRevB.98.220303}.
\newblock URL \url{https://link.aps.org/doi/10.1103/PhysRevB.98.220303}.

\bibitem[Rakovszky et~al.(2018)Rakovszky, Pollmann, and von
  Keyserlingk]{Rakovsky2018}
Tibor Rakovszky, Frank Pollmann, and C.~W. von Keyserlingk.
\newblock Diffusive hydrodynamics of out-of-time-ordered correlators with
  charge conservation.
\newblock \emph{Phys. Rev. X}, 8:\penalty0 031058, Sep 2018.
\newblock \doi{10.1103/PhysRevX.8.031058}.
\newblock URL \url{https://link.aps.org/doi/10.1103/PhysRevX.8.031058}.

\bibitem[Susskind(2016)]{Susskind2016}
Leonard Susskind.
\newblock Computational complexity and black hole horizons.
\newblock \emph{Fortschritte der Physik}, 64\penalty0 (1):\penalty0 24--43,
  2016.
\newblock \doi{https://doi.org/10.1002/prop.201500092}.
\newblock URL
  \url{https://onlinelibrary.wiley.com/doi/abs/10.1002/prop.201500092}.

\bibitem[Brown et~al.(2016{\natexlab{a}})Brown, Roberts, Susskind, Swingle, and
  Zhao]{Brown16}
Adam~R. Brown, Daniel~A. Roberts, Leonard Susskind, Brian Swingle, and Ying
  Zhao.
\newblock Holographic complexity equals bulk action?
\newblock \emph{Phys. Rev. Lett.}, 116:\penalty0 191301, May
  2016{\natexlab{a}}.
\newblock \doi{10.1103/PhysRevLett.116.191301}.
\newblock URL \url{https://link.aps.org/doi/10.1103/PhysRevLett.116.191301}.

\bibitem[Brown et~al.(2016{\natexlab{b}})Brown, Roberts, Susskind, Swingle, and
  Zhao]{Brown16prd}
Adam~R. Brown, Daniel~A. Roberts, Leonard Susskind, Brian Swingle, and Ying
  Zhao.
\newblock Complexity, action, and black holes.
\newblock \emph{Phys. Rev. D}, 93:\penalty0 086006, Apr 2016{\natexlab{b}}.
\newblock \doi{10.1103/PhysRevD.93.086006}.
\newblock URL \url{https://link.aps.org/doi/10.1103/PhysRevD.93.086006}.

\bibitem[Chapman et~al.(2018)Chapman, Heller, Marrochio, and
  Pastawski]{Chapman18}
Shira Chapman, Michal~P. Heller, Hugo Marrochio, and Fernando Pastawski.
\newblock Toward a definition of complexity for quantum field theory states.
\newblock \emph{Phys. Rev. Lett.}, 120:\penalty0 121602, Mar 2018.
\newblock \doi{10.1103/PhysRevLett.120.121602}.
\newblock URL \url{https://link.aps.org/doi/10.1103/PhysRevLett.120.121602}.

\bibitem[Molina-Vilaplana and del Campo(2018)]{Molina-Vilaplana18}
J.~Molina-Vilaplana and A.~del Campo.
\newblock Complexity functionals and complexity growth limits in continuous
  mera circuits.
\newblock \emph{Journal of High Energy Physics}, 2018\penalty0 (8):\penalty0
  12, Aug 2018.
\newblock ISSN 1029-8479.
\newblock \doi{10.1007/JHEP08(2018)012}.
\newblock URL \url{https://doi.org/10.1007/JHEP08(2018)012}.

\bibitem[H{\"o}rnedal et~al.(2022)H{\"o}rnedal, Carabba, Matsoukas-Roubeas, and
  del Campo]{Hornedal2022}
Niklas H{\"o}rnedal, Nicoletta Carabba, Apollonas~S. Matsoukas-Roubeas, and
  Adolfo del Campo.
\newblock Ultimate speed limits to the growth of operator complexity.
\newblock \emph{Communications Physics}, 5\penalty0 (1):\penalty0 207, Aug
  2022.
\newblock ISSN 2399-3650.
\newblock \doi{10.1038/s42005-022-00985-1}.
\newblock URL \url{https://doi.org/10.1038/s42005-022-00985-1}.

\bibitem[Parker et~al.(2019)Parker, Cao, Avdoshkin, Scaffidi, and
  Altman]{Parker2019}
Daniel~E. Parker, Xiangyu Cao, Alexander Avdoshkin, Thomas Scaffidi, and Ehud
  Altman.
\newblock A universal operator growth hypothesis.
\newblock \emph{Phys. Rev. X}, 9:\penalty0 041017, Oct 2019.
\newblock \doi{10.1103/PhysRevX.9.041017}.
\newblock URL \url{https://link.aps.org/doi/10.1103/PhysRevX.9.041017}.

\bibitem[Barb{\'o}n et~al.(2019)Barb{\'o}n, Rabinovici, Shir, and
  Sinha]{Barbon2019}
J.L.F. Barb{\'o}n, E.~Rabinovici, R.~Shir, and R.~Sinha.
\newblock On the evolution of operator complexity beyond scrambling.
\newblock \emph{J. High Energ. Phys.}, 2019\penalty0 (10):\penalty0 264,
  October 2019.
\newblock ISSN 1029-8479.
\newblock \doi{10.1007/JHEP10(2019)264}.
\newblock URL \url{https://doi.org/10.1007/JHEP10(2019)264}.

\bibitem[Rabinovici et~al.(2021)Rabinovici, S{\'a}nchez-Garrido, Shir, and
  Sonner]{Rabinovici2021}
E.~Rabinovici, A.~S{\'a}nchez-Garrido, R.~Shir, and J.~Sonner.
\newblock Operator complexity: a journey to the edge of {Krylov} space.
\newblock \emph{J. High Energ. Phys.}, 2021\penalty0 (6):\penalty0 62, June
  2021.
\newblock ISSN 1029-8479.
\newblock \doi{10.1007/JHEP06(2021)062}.
\newblock URL \url{https://doi.org/10.1007/JHEP06(2021)062}.

\bibitem[Caputa et~al.(2021)Caputa, Magan, and Patramanis]{Caputa2021}
Pawel Caputa, Javier~M. Magan, and Dimitrios Patramanis.
\newblock Geometry of {Krylov} {Complexity}.
\newblock \emph{arXiv:2109.03824}, September 2021.
\newblock URL \url{http://arxiv.org/abs/2109.03824}.

\bibitem[Kubo(1957)]{Kubo1957}
Ryogo Kubo.
\newblock Statistical-mechanical theory of irreversible processes. i. general
  theory and simple applications to magnetic and conduction problems.
\newblock \emph{Journal of the Physical Society of Japan}, 12\penalty0
  (6):\penalty0 570--586, 1957.
\newblock \doi{10.1143/JPSJ.12.570}.
\newblock URL \url{https://doi.org/10.1143/JPSJ.12.570}.

\bibitem[Ness et~al.(2021)Ness, Lam, Alt, Meschede, Sagi, and
  Alberti]{Ness2021}
Gal Ness, Manolo~R. Lam, Wolfgang Alt, Dieter Meschede, Yoav Sagi, and Andrea
  Alberti.
\newblock Observing crossover between quantum speed limits.
\newblock \emph{Science Advances}, 7\penalty0 (52):\penalty0 eabj9119, 2021.
\newblock \doi{10.1126/sciadv.abj9119}.
\newblock URL \url{https://www.science.org/doi/abs/10.1126/sciadv.abj9119}.

\bibitem[Hauke et~al.(2016)Hauke, Heyl, Tagliacozzo, and Zoller]{Hauke2016}
Philipp Hauke, Markus Heyl, Luca Tagliacozzo, and Peter Zoller.
\newblock Measuring multipartite entanglement through dynamic susceptibilities.
\newblock \emph{Nature Physics}, 12\penalty0 (8):\penalty0 778--782, 2016.
\newblock \doi{10.1038/nphys3700}.
\newblock URL \url{https://doi.org/10.1038/nphys3700}.

\bibitem[Wang et~al.(2009)Wang, Sun, and Wang]{Wang09}
Xiaoguang Wang, Zhe Sun, and Z.~D. Wang.
\newblock Operator fidelity susceptibility: An indicator of quantum
  criticality.
\newblock \emph{Phys. Rev. A}, 79:\penalty0 012105, Jan 2009.
\newblock \doi{10.1103/PhysRevA.79.012105}.
\newblock URL \url{https://link.aps.org/doi/10.1103/PhysRevA.79.012105}.

\bibitem[Andersson(2019)]{Andersson2019}
Ole Andersson.
\newblock \emph{Holonomy in Quantum Information Geometry}.
\newblock PhD thesis, Stockholm University, 2019.

\bibitem[Ness et~al.(2022)Ness, Alberti, and Sagi]{Ness2022}
Gal Ness, Andrea Alberti, and Yoav Sagi.
\newblock Quantum speed limit for states with a bounded energy spectrum.
\newblock \emph{Phys. Rev. Lett.}, 129:\penalty0 140403, Sep 2022.
\newblock \doi{10.1103/PhysRevLett.129.140403}.
\newblock URL \url{https://link.aps.org/doi/10.1103/PhysRevLett.129.140403}.

\bibitem[Levitin and Toffoli(2009)]{Levitin2009}
Lev~B. Levitin and Tommaso Toffoli.
\newblock Fundamental limit on the rate of quantum dynamics: The unified bound
  is tight.
\newblock \emph{Phys. Rev. Lett.}, 103:\penalty0 160502, Oct 2009.
\newblock \doi{10.1103/PhysRevLett.103.160502}.
\newblock URL \url{https://link.aps.org/doi/10.1103/PhysRevLett.103.160502}.

\bibitem[Dymarsky and Smolkin(2021)]{Dymarsky2021}
Anatoly Dymarsky and Michael Smolkin.
\newblock Krylov complexity in conformal field theory.
\newblock \emph{Phys. Rev. D}, 104:\penalty0 L081702, Oct 2021.
\newblock \doi{10.1103/PhysRevD.104.L081702}.
\newblock URL \url{https://link.aps.org/doi/10.1103/PhysRevD.104.L081702}.

\bibitem[Alhambra et~al.(2020)Alhambra, Riddell, and
  Garc\'{\i}a-Pintos]{Alhambra2020}
\'Alvaro~M. Alhambra, Jonathon Riddell, and Luis~Pedro Garc\'{\i}a-Pintos.
\newblock Time evolution of correlation functions in quantum many-body systems.
\newblock \emph{Phys. Rev. Lett.}, 124:\penalty0 110605, Mar 2020.
\newblock \doi{10.1103/PhysRevLett.124.110605}.
\newblock URL \url{https://link.aps.org/doi/10.1103/PhysRevLett.124.110605}.

\bibitem[Tuckerman(2010)]{Tuckerman}
Mark~E. Tuckerman.
\newblock \emph{Statistical Mechanics: Theory and Molecular Simulation}.
\newblock Oxford University Press, 2010.
\newblock \doi{https://doi.org/10.1002/anie.201105752}.

\bibitem[Ueda(2010)]{Ueda}
Masahito Ueda.
\newblock \emph{Fundamentals and New Frontiers of Bose-Einstein Condensation}.
\newblock WORLD SCIENTIFIC, 2010.
\newblock \doi{10.1142/7216}.
\newblock URL \url{https://www.worldscientific.com/doi/abs/10.1142/7216}.

\bibitem[Mazenko(2006)]{Mazenko}
Gene~F. Mazenko.
\newblock \emph{Nonequilibrium Statistical Mechanics}.
\newblock John Wiley Sons, 2006.
\newblock ISBN 9783527618958.
\newblock \doi{https://doi.org/10.1002/9783527618958}.

\bibitem[Pake(1962)]{Pake}
G.E. Pake.
\newblock \emph{Paramagnetic Resonance: An Introductory Monograph}.
\newblock Number v. 1 in Frontiers in physics. W.A. Benjamin, 1962.
\newblock URL \url{https://books.google.lu/books?id=B8pEAAAAIAAJ}.

\bibitem[Brenes et~al.(2020)Brenes, Pappalardi, Goold, and Silva]{Brenes20}
Marlon Brenes, Silvia Pappalardi, John Goold, and Alessandro Silva.
\newblock Multipartite entanglement structure in the eigenstate thermalization
  hypothesis.
\newblock \emph{Phys. Rev. Lett.}, 124:\penalty0 040605, Jan 2020.
\newblock \doi{10.1103/PhysRevLett.124.040605}.
\newblock URL \url{https://link.aps.org/doi/10.1103/PhysRevLett.124.040605}.

\bibitem[Braunstein et~al.(1996)Braunstein, Caves, and Milburn]{Braunstein96}
Samuel~L. Braunstein, Carlton~M. Caves, and G.J. Milburn.
\newblock Generalized uncertainty relations: Theory, examples, and lorentz
  invariance.
\newblock \emph{Annals of Physics}, 247\penalty0 (1):\penalty0 135--173, 1996.
\newblock ISSN 0003-4916.
\newblock \doi{https://doi.org/10.1006/aphy.1996.0040}.
\newblock URL
  \url{https://www.sciencedirect.com/science/article/pii/S0003491696900408}.

\bibitem[Giovannetti et~al.(2003)Giovannetti, Lloyd, and
  Maccone]{Giovannetti2003}
Vittorio Giovannetti, Seth Lloyd, and Lorenzo Maccone.
\newblock Quantum limits to dynamical evolution.
\newblock \emph{Phys. Rev. A}, 67:\penalty0 052109, May 2003.
\newblock \doi{10.1103/PhysRevA.67.052109}.
\newblock URL \url{https://link.aps.org/doi/10.1103/PhysRevA.67.052109}.

\bibitem[Giovannetti et~al.(2004)Giovannetti, Lloyd, and
  Maccone]{Giovannetti2004}
Vittorio Giovannetti, Seth Lloyd, and Lorenzo Maccone.
\newblock The speed limit of quantum unitary evolution.
\newblock \emph{Journal of Optics B: Quantum and Semiclassical Optics},
  6\penalty0 (8):\penalty0 S807--S810, jul 2004.
\newblock \doi{10.1088/1464-4266/6/8/028}.
\newblock URL \url{https://doi.org/10.1088/1464-4266/6/8/028}.

\bibitem[del Campo et~al.(2017)del Campo, Molina-Vilaplana, and
  Sonner]{Delcampo17}
A.~del Campo, J.~Molina-Vilaplana, and J.~Sonner.
\newblock Scrambling the spectral form factor: Unitarity constraints and exact
  results.
\newblock \emph{Phys. Rev. D}, 95:\penalty0 126008, Jun 2017.
\newblock \doi{10.1103/PhysRevD.95.126008}.
\newblock URL \url{https://link.aps.org/doi/10.1103/PhysRevD.95.126008}.

\bibitem[Xu et~al.(2021)Xu, Chenu, Prosen, and del Campo]{Xu21}
Zhenyu Xu, Aurelia Chenu, Toma\ifmmode\check{z}\else\v{z}\fi{} Prosen, and
  Adolfo del Campo.
\newblock Thermofield dynamics: Quantum chaos versus decoherence.
\newblock \emph{Phys. Rev. B}, 103:\penalty0 064309, Feb 2021.
\newblock \doi{10.1103/PhysRevB.103.064309}.
\newblock URL \url{https://link.aps.org/doi/10.1103/PhysRevB.103.064309}.

\bibitem[Okuyama and Ohzeki(2018{\natexlab{c}})]{Okuyama2018Comment}
Manaka Okuyama and Masayuki Ohzeki.
\newblock Comment on ‘energy-time uncertainty relation for driven quantum
  systems’.
\newblock \emph{Journal of Physics A: Mathematical and Theoretical},
  51\penalty0 (31):\penalty0 318001, jun 2018{\natexlab{c}}.
\newblock \doi{10.1088/1751-8121/aacb90}.
\newblock URL \url{https://dx.doi.org/10.1088/1751-8121/aacb90}.

\end{thebibliography}

\end{document}